\documentclass{pasa}%
\usepackage[pdftex]{graphicx}
\usepackage{aas_macros}
\usepackage{multirow,comment,multicol}
\usepackage[titletoc,title]{appendix}
\usepackage[usenames,dvipsnames,svgnames,hyperref]{xcolor}
\usepackage{varwidth}
\usepackage{tikz}
\usetikzlibrary{positioning,calc,backgrounds}
\tikzstyle{line}=[draw]
\tikzstyle{arrow}=[draw, -latex] 
\usepackage{tcolorbox}
\usepackage[
    pdftex,
    pdfpagelabels,
    breaklinks=true,
    bookmarks=true,
    bookmarksopen=false,
    bookmarksopenlevel=2
    bookmarksnumbered=true,
    bookmarkstype=toc,
    colorlinks=true,
    citecolor=RoyalBlue,
    linkcolor=ForestGreen,
    menucolor=Teal,
    urlcolor=DarkOrange,
]{hyperref}
\usepackage{subcaption}
\usepackage{amsfonts}
\usepackage{longtable}

\newcommand{\Eqref}[1]{Eq.~(\ref{#1})}
\newcommand{\Figref}[1]{Fig.~\ref{#1}}
\newcommand{\Secref}[1]{\S\ref{#1}}  
\newcommand{\Tableref}[1]{Table~\ref{#1}}

\def \R{\mathcal{R}}
\def \ELL{\mathcal{L}}
\def \Nse{N_{\rm se}}
\def \Nv{N_{\rm v}}
\def \Ncell{N_{\rm cell}}
\def \ELLth{\ELL_{\rm th}}
\def \Vr{\mathcal{V}_{\rm r}}
\def \Thetaop{\Theta_{\rm op}}
\def \ellx{\ell_{\rm x}}
\def \ellv{\ell_{\rm v}}

\def \Mpch{h^{-1}{\rm Mpc}}
\def \kpch{h^{-1}{\rm kpc}}
\def \Msunh{h^{-1}{\rm M}_\odot}

\newcommand*{\ditto}{---\texttt{"}---}

\title[\textsc{VELOCIraptor}]{Hunting for Galaxies and Halos in simulations with \textsc{VELOCIraptor}}

\author[Elahi et al.]{
Pascal J. Elahi$^{1,2,\dagger}$, 
Rodrigo Ca\~nas$^{1,2}$, 
Rhys J.J. Poulton$^{1,2}$, 
Rodrigo J. Tobar$^{1}$, 
James S. Willis$^{3}$,
Claudia del P. Lagos$^{1,2}$, 
Chris Power$^{1,2}$,
Aaron S.~G. Robotham$^{1}$,
\affil{
$^1$International Centre for Radio Astronomy Research, University of Western Australia, 35 Stirling Highway, Crawley, WA 6009, Australia}
\affil{
$^2$ARC Centre of Excellence for All Sky Astrophysics in 3 Dimensions (ASTRO 3D)
}
\affil{
$^3$Institute for Computational Cosmology (ICC),
}%
\affil{$^\dagger$Email: pascal.elahi@icrar.org}
}%

\jid{PASA}
\doi{10.1017/pas.\the\year.xxx}
\jyear{\the\year}

\begin{document}

\begin{frontmatter}
\maketitle

\begin{abstract}
We present \textsc{VELOCIraptor},  a massively parallel \textit{galaxy/(sub)halo finder} that is also capable of robustly identifying tidally disrupted objects and separate stellar halos from galaxies. The code is written in \textsc{c++11}, use the MPI and OpenMP API's for parallelisation, and includes python tools to read/manipulate the data products produced.  We demonstrate the power of the \textsc{VELOCIraptor} (sub)halo finder, showing how it can identify subhalos deep within the host that have negligible density contrasts to their parent halo. We find a subhalo mass-radial distance dependence: large subhalos with mass ratios of $\gtrsim10^{-2}$ are more common in the central regions that smaller subhalos, a result of dynamical friction and low tidal mass loss rates. This dependence is completely absent in (sub)halo finders in common use, which generally search for substructure in configuration space, yet is present in codes that track particles belonging to halos as they fall into other halos, such as \textsc{hbt+}. \textsc{VELOCIraptor} largely reproduces the dependence seen without tracking, finding a similar radial dependence to \textsc{hbt+} in well resolved halos from our limited resolution fiducial simulation. 
\end{abstract}

\end{frontmatter}

\section{Introduction}
\label{sec:intro}
Running a cosmological simulation, whether N-Body or full hydrodynamical, is the first step in understanding cosmic structure formation and the evolution of galaxies. A critical step in extracting information from sophisticated, multi-billion particle simulations is the identification of structures, like dark matter halos and synthetic galaxies. Identifying (sub)structures is a non-trivial task and has led to the development of equally sophisticated structure finders \cite[see][for an overview of (sub)halo/galaxy finding]{knebe2011,onions2012,knebe2012a,knebe2013a}. A variety of codes exist that attempt to excise structures of interest from simulations, with most focusing on searching for overdense, gravitationally self-bound regions within cosmological simulations. For cosmological N-body simulations, these objects are dark matter halos, for hydrodynamical simulations these objects can be galaxies. 

\par
The two most common pure halo finders are Friends-of-Friends algorithms \cite[e.g.][]{fof} and Spherical Overdensity algorithms \cite[e.g.][]{laceycole1994}, the former using a linking length based on a desired density criterion, the latter identifying density peaks and grouping all particles within a spherical region that encloses some density \cite[see][for a more thorough discussion and comparison of halo finding]{knebe2011}.

\par 
Beyond halo finders are those that also attempted to excise substructures residing within the gravitationally collapsed, nonlinear environment of halos, so-called subhalo finders. Subhalo finders can be broadly classified into two types: configuration-space finders and phase-space finders. Older, more common configuration-space finders, like {\sc ahf} \cite[][]{ahf}, {\sc subfind} \cite[][]{subfind}, {\sc adaptahop} \cite[][]{tweed2009a}, search for physical overdensities or clustering in configuration space\footnote{In practice, even configuration-space finders are pseudo phase-space finders as candidate objects must be passed through an unbinding process, whereby unbound particles are removed from a candidate, to return sensible results.}. Phase-space finders, like {\sc hsf} \cite[][]{hsf} and {\sc rockstar} \cite[][]{rockstar}, use extra velocity information to identify overdensities and clustering in the full phase-space. 

\par 
Different (sub)halo finders suffer from different issues \cite[see][for a in depth discussion of structure finding]{knebe2013a}. Configuration-space base finders rely on saddle points in the density field in some form or another to separate structures. Consequently, subhalos are artificially truncated as they fall towards pericentre and grow again as the move out to apocentre \cite[see][for specific examples using \textsc{subfind} \& \textsc{ahf}]{muldrew2011,behroozi2015a}. Phase-space finders are better able to separate these structures since they will overlap less in phase-space, and in principle need not inherently shrink/grow the mass associated with subhalos as they move towards pericentre/apocentre. 

\par 
Here we present \textsc{VELOCIraptor} \cite[formerly known as STructure Finder, {\sc stf}][]{elahi2011}, a phase-space (sub)halo finder capable of identifying dark matter halos and galaxies\footnote{Freely available \href{https://github.com/pelahi/VELOCIraptor-STF.git}{\url{https://github.com/pelahi/VELOCIraptor-STF.git}}. Documentation is found at \href{http://velociraptor-stf.readthedocs.io/en/latest/}{\url{http://velociraptor-stf.readthedocs.io/en/latest/}}}. This code can ingest both pure N-Body simulation input and hydrodynamical data. Here we present significant update to the original algorithm described in \cite{elahi2011}. 

\par
Our paper is organised as follows: in section \Secref{sec:velociraptor}, we outline the code package, present tests of our algorithm in \Secref{sec:results:velociraptor} and conclude in \Secref{sec:discussion} with a summary and discussion.

\section{Identifying Structures with \textsc{VELOCIraptor}}
\label{sec:velociraptor}
\textsc{VELOCIraptor} is a (sub)halo finder that identifies structures in a multi-stage process, the exact details depending on the operational mode it is being used in: identifying dark matter halos, dark matter halos+baryonic content, or just galaxies. \textsc{VELOCIraptor} is built on \textsc{STF} \cite[][]{elahi2011}, providing significant upgrades to the halo finding algorithm, treatment of baryons, the mass reconstruction of major merger events, along with parallelisation and integration into N-body codes \cite[specifically \textsc{swift}][]{swiftsimcode}. We describe the various aspects of our code below. For readers interested in input interfaces, output, and general modes of operation we suggest skipping to \Secref{sec:velociraptor:summary}. Readers interested in the main benefits and results of \textsc{VELOCIraptor} can skip to \Secref{sec:results:velociraptor}. 

\par
\begin{figure}
\centering
\begin{tikzpicture}[
    roundnode/.style={rectangle, rounded corners = 5pt, draw=ForestGreen, very thick, minimum size=5mm,align=center},
    squarednode/.style={rectangle, draw=Cerulean, very thick, text width=0.4\textwidth, align=left},
    node distance=5mm
    ]
    \node[squarednode](config){
    {\bf Setup}: 
    Read configuration options.
    };
    \node[squarednode](input)[below=of config]{
    {\bf Input}: 
    Read input file \& setup MPI decomposition if required.
    };
    \node[squarednode](field)[below=of input]{
    {\bf Field Search}: 
    Each MPI process searches for field halos, linking across MPI domains and localising halos to single MPI process.
    };
    \node[squarednode](subs)[below=of field]{
    {\bf Substructure/Merger Search}: 
    Search for substructures/mergers in all (sub)structures large enough to have substructures above the minimum number of particles.
    };
    \node[squarednode](props)[below=of subs]{
    {\bf Analysis}: 
    Calculate properties for all (sub)halos.
    };
    \node[squarednode](output)[below=of props]{
    {\bf Output}: 
    Each MPI process outputs files containing bulk properties and the particle IDs of all particles within halos.
    };
    
    \draw[thick,->] (config.south) -- (input.north);
    \draw[thick,->] (input.south) -- (field.north);
    \draw[thick,->] (field.south) -- (subs.north);
    \draw[thick,->] (subs.south) -- (props.north);
    \draw[thick,->] (props.south) -- (output.north);

\end{tikzpicture}
    \caption{Activity chart of \textsc{VELOCIraptor}.}
    \label{fig:velociraptor:scheme}
\end{figure}
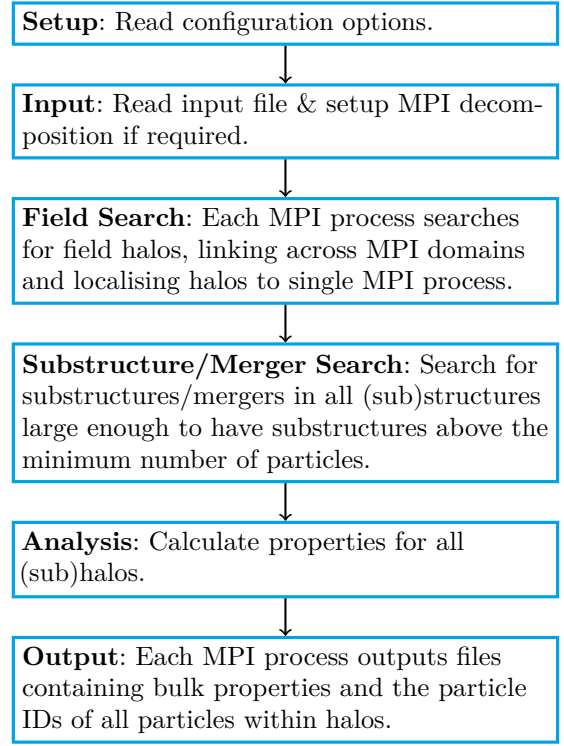

 
The identification process proceeds in a two stage approach: 1) identify field halos/galaxies; 2) for each field object search for substructure using phase-space information. Unlike almost all other structure finders currently available, this algorithm is also capable of robustly identifying tidally disrupted objects \cite[see][]{elahi2013a} along with self-bound, physically dense halos/galaxies. A flow chart describing the operational stages is shown in \Figref{fig:velociraptor:scheme}. 

\subsection{Field Halos}
\label{sec:velociraptor:halos}
The code first identifies candidate halos using a 3DFOF algorithm \cite[3D Friends-of-Friends in configuration space, see][]{fof}, linking particles together if 
\begin{gather}
    \frac{\left({\bf x}_i-{\bf x}_j\right)^2}{\ellx^2}<1,\label{eqn:fof3d}
\end{gather}
where ${\bf x}_i$ is the $i^{\rm th}$ particle's position, and $\ellx$ is the linking length. This initial linking can also make use of a particle's type, whether dark matter (N-body), or gas, star (baryon). Cosmological simulations typically set $\ellx=0.2$ times the inter-particle spacing. 

\par 
Simple FOF algorithms are susceptible to artificially joining two structures together by a single (or a few) particle(s), a so-called particle bridge. We appeal to the physics of the structures we seek to identify, i.e., virialised halos, and use velocity information\footnote{In general, artificial particle bridges could be removed by identifying a particle(s) that, if removed,  would split the structure into several structures, i.e., those particles that have groups of links whose sole common link is the particle itself.}. For each structure $k$ we calculate a velocity dispersion, $\sigma_{v,k}$, and apply a 6DFOF, 
\begin{equation}
    \frac{\left({\bf x}_i-{\bf x}_j\right)^2}{\ellx^2}+\frac{\left({\bf v}_i-{\bf v}_j\right)^2}{\ellv^2}<1,\label{eqn:fof6d}
\end{equation}
which splits virialised structures connected by dynamically unrelated particle bridges and tends to remove very unbound particles that may have been grouped by the original FOF algorithm. Here $\ellv=\alpha_{\rm v}\sigma_{v,k}$, and $\alpha_{\rm v}$ is a scaling term on the order of unity. 

\paragraph{Addition of Baryons:}
Simulations can contain both N-body (Dark Matter, DM) particles and other particle types and along with the inclusion of extra forces, like the addition of gas tracers and hydrodynamical forces. Fully hydrodynamical cosmological simulations often contain either gas particles (or tracers for codes such as AREPO \citealp{arepo}, or cells such as RAMSES \citealp{ramses}), star particles, and even sink particles representing supermassive black holes. These baryons tracers can be treated in a special fashion by VELOCIraptor if the appropriate flags are set. If desired, specific particle types can be searched, such as stars to produce a galaxy catalogue. The code can also search all particle types, either treating all particles equally or allowing for special linking behaviour dependent on particle type. 

\par 
The two most common modes of operation are either to assign baryonic particles to DM structures, so-called \textit{DM+Baryons}, or to search only star particles and identify galaxies, so-called \textit{Galaxies+Baryons}. We discuss how the field search operates in both these modes. 

\par
Since gas particles are subject to hydrodynamical forces and can clump together to form long filaments, applying a simple FOF algorithm can leading to the artificial linking together of several dynamically distinct structures. Hence the typical mode of operation to group both DM and baryons together is to produce FOF links using DM particles only, i.e., a DM particle can link to other DM particles and baryon particles, but baryon particles are ignored when searching for new FOF links. An application of this mode has been applied to hydrodynamical zoom simulations \cite[e.g.][]{nifty3,nifty5}.

\par 
When searching for galaxies using star particles, we first identify 3DFOF stellar structures. These structures are then searched using a 6DFOF, with the critical difference between the DM search being that we keep track of the star particles linked in the 3DFOF but not linked in the 6DFOF as a structure. This remnant 3DFOF represents the diffuse, kinematically distinct stellar halos that surround galaxies. An application of this mode has been applied to hydrodynamical  simulations to look at the sizes of galaxies and a preliminary investigation of diffuse stellar halos \cite[][Canas et al, in prep]{canas2018a}. The code can also use star particles as a basis for links to assign other baryonic particle types to structures in a similar fashion to the DM mode described above. 

\subsection{Subhalos \& Streams}
\label{sec:velociraptor:substructures}
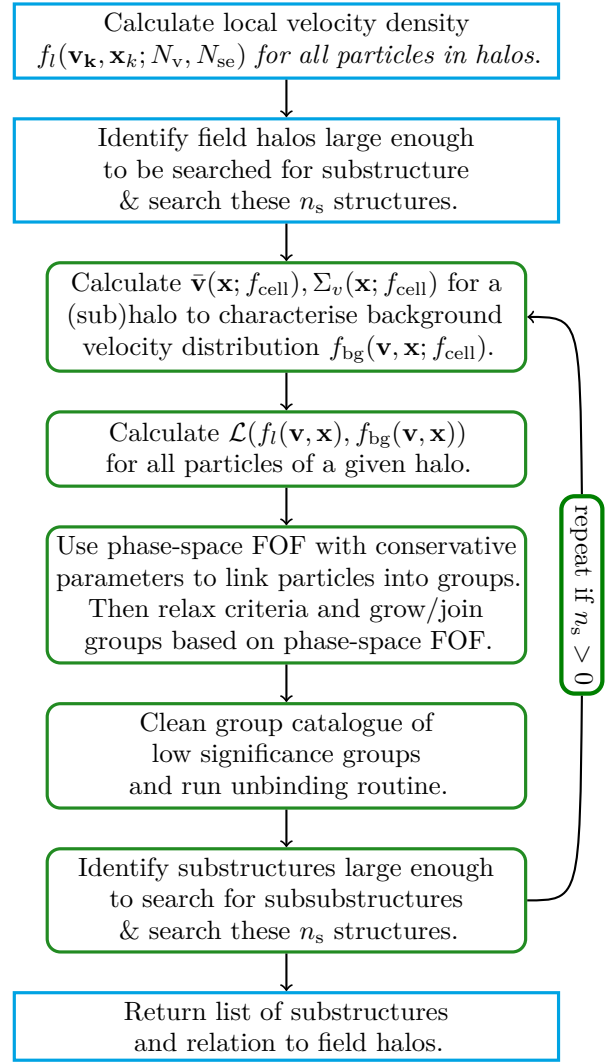
\begin{figure}
\centering
\begin{tikzpicture}[
    roundnode/.style={rectangle, rounded corners = 5pt, draw=ForestGreen, very thick, text width=0.35\textwidth,align=center},
    squarednode/.style={rectangle, draw=Cerulean, very thick, minimum size=5mm, text width=0.4\textwidth, align=center},
    node distance=5mm
    ]
    \node[squarednode](fl){
    Calculate local velocity density $f_l({\bf {v}_k},{\bf x}_k;\Nv,\Nse)$ {\em for all particles in halos}.
    };
    \node[squarednode](searching)[below=of fl]{
    Identify field halos large enough to be searched for substructure \& search these $n_{\rm s}$ structures.
    };
    \node[roundnode](fb)[below=of searching]{
    Calculate $\bar{\bf v}({\bf x};f_{\rm cell}),\Sigma_v({\bf x};f_{\rm cell})$ for a (sub)halo to characterise background velocity distribution $f_{\rm bg}({\bf v}, {\bf x};f_{\rm cell})$.
    };
    \node[roundnode](outliers)[below=of fb]{
    Calculate $\ELL(f_l({\bf v},{\bf x}),f_{\rm bg}({\bf v},{\bf x}))$ for all particles of a given halo.
    };
    \node[roundnode](link)[below=of outliers]{
    Use phase-space FOF with conservative parameters to link particles into groups. Then relax criteria and grow/join groups based on phase-space FOF.
    };
    \node[roundnode](clean)[below=of link]{
    Clean group catalogue of low significance groups and run unbinding routine.
    };
    \node[roundnode](newsearch)[below=of clean]{
    Identify substructures large enough to search for subsubstructures \& search these $n_{\rm s}$ structures.
    };
    \node[rectangle,  rounded corners = 5pt, draw=Green, ultra thick, rotate=-90, yshift=15pt](loop)[right=of link, anchor=north]{
    \begin{varwidth}{0.2\textwidth}
    repeat if $n_{\rm s}>0$
    \end{varwidth}
    };
    \node[squarednode](result)[below=of newsearch]{
    Return list of substructures and relation to field halos.
    };
    
    \draw[thick,->] (fl.south) -- (searching.north);
    \draw[thick,->] (searching.south) -- (fb.north);
    \draw[->,thick] (fb.south) -- (outliers.north);
    \draw[->,thick] (outliers.south) -- (link.north);
    \draw[->,thick] (link.south) -- (clean.north);
    \draw[->,thick] (clean.south) -- (newsearch.north);
    \draw[-,thick] (newsearch.east) .. controls +(right:7mm) .. (loop.east);
    \draw[<-,thick] (fb.east) .. controls +(right:7mm) .. (loop.west);
    \draw[->,thick] (newsearch.south) -- (result.north);
\end{tikzpicture}
    \caption{Activity chart for identifying substructures.}
    \label{fig:substructure:scheme}
\end{figure}

We briefly describe the specifics of identifying substructures here as it is discussed in \cite{elahi2011}. Substructures are identified using a phase-space FOF algorithm on particles that appear to be {\em dynamically distinct} from the mean ``Maxwellian'' halo background, i.e., particles which have a local velocity distribution that differs significantly from the mean, smooth background halo. This approach is capable of not only finding subhalos, but also tidal debris surrounding subhalos as well as tidal streams from completely disrupted subhalos. The method for identifying substructure is shown in \Figref{fig:substructure:scheme}.

\paragraph{Dynamically distinct particles:}
\label{sec:velociraptor:substructures:particles}
The algorithm identifies particles that are dynamically distinct from a background distribution by examining velocity space assuming that that a halo's velocity distribution can be split into a virialised background and substructures. To illustrate this method, consider the phase-space distribution function:
\begin{align}
  F({\bf x},{\bf v})=\rho({\bf x})f({\bf v}).
\end{align}
Here we assume the distribution function is separable into $\rho({\bf x})$ and $f({\bf v})$, the physical and velocity density distribution functions respectively. Assuming Gaussian velocity distributions for a substructure and a halo, the distribution ratio of a substructure S to the background bg at a given $({\bf x},{\bf v})$ is:
\begin{align} 
  \frac{F_{\rm S}({\bf x},{\bf v})}{F_{\rm bg}({\bf x},{\bf v})}=
  {
  \left[\frac{\rho_{\rm S}({\bf x})}{\rho_{\rm bg}({\bf x})}\right]
  }
  {
  \left[\frac{\sigma_{\rm bg}^3}{\sigma_{\rm S}^3}\right]
  }
  {
  \left[\frac{{\rm e}^{-({\bf v}-{\bf v}_{\rm S})^2/2\sigma^2_{\rm S}}}{{\rm e}^{-({\bf v}-{\bf v}_{\rm bg})^2/2\sigma^2_{\rm bg}}}\right]
  }
  .
  \label{eqn:distribratio}
\end{align}
This ratio has three terms: { the physical density contrast}; { velocity dispersion contrast}; { and a ratio of Gaussian terms}. Subhalos are dynamically cold overdensities, unlike tidal streams, which can have negligible density contrasts and velocity dispersion comparable with the background. Hence, it is common practice to focus on the density ratio to identify subhalos. However, regardless of whether a substructure is a subhalo or tidal debris, the velocity distribution of the particles belonging to the substructure will differ from the background. These particles will have a ratio of at least ${\exp(\delta v^2/2\sigma_{\rm bg}^2)}$. 

\par 
This exponential factor, a measure of orbit clustering, is key to our algorithm. Instead of estimating the full phase-space density at a particle's phase-space position ${\bf X}$, we measure {\em local velocity density}, $f_{\rm l}({\bf v}|{\bf x})$, as this is less computationally expense and not as noisy. We then divide out the expected velocity density of the background, $f_{\rm bg}({\bf v}|{\bf x})$, neglecting the first term in \Eqref{eqn:distribratio} at this stage. Particles belonging to velocity distributions that differ from the background will have ratios of $f_{\rm l}/f_{\rm bg}\gg1$.

\par 
The {\em local} velocity density of a particle $k$, $f_{\rm l}({\bf v}_k)$, is measured using a kernel-scheme with an Epanechnikov smoothing kernel \cite[][]{enbid}. This density is calculated using $\Nv$ nearest velocity neighbours from the set of $\Nse$ nearest physical neighbours, where $\Nv\leq\Nse$\footnote{Using a subset of physical neighbours to measure the local velocity density will give a biased result but as the goal is to highlight any clustering in velocity space, this is perfectly acceptable}. Typical values are $\Nv=32,\Nse=256$. 

\par
The mean background velocity density is characterised by a multivariate Gaussian\footnote{Numerical simulations show the velocity distribution of a small region of a cosmological halo are reasonably characterised by a multivariate Gaussian \cite[e.g.][]{vogelsberger2009b}.}, thus, the expected {\em background} velocity density for a particle $k$ with velocity ${\bf v}_k$ is
\begin{align}
    f_{\rm bg}({\bf v}_k)=\frac{\exp\left[-\tfrac{1}{2}({\bf v}_k-\bar{\bf v}({\bf x}_k))\Sigma^{-1}_v({\bf x}_k)({\bf v}_k-\bar{\bf v}({\bf x}_k))\right]}
    {(2\pi)^{3/2}|\Sigma_v({\bf x}_k)|^{1/2}},
    \label{eqn:fvbg}
\end{align}
where $\bar{\bf v}$ is the mean velocity, and $\Sigma_v$ is the matrix representation of velocity dispersion tensor about $\bar{\bf v}$, both of which depend on the position within the halo, ${\bf x}$.

\par 
The mean field is estimated by splitting the halo into volumes containing enough particles so that the statistical error on bulk quantities calculated for a cell is negligible but not so large that density (and thus the velocity dispersion) varies greatly across the volume. 
To balance these competing effects, we split the halo into cells containing $\Ncell$ particles using a KD-Tree \citep{kdtree,appel1985,treecode}, iteratively splitting along the spatial dimension that maximises Shannon entropy, $S$. We calculate $S$ for each dimension by binning particles in $n_{\rm bins}$ that span the extent of the dimension using the formula
\begin{align}
    S=\frac{1}{\log n_{\rm bins}}\sum\limits_k^{n_{\rm bins}}-\frac{m_k}{m_{\rm tot}}\log\frac{m_k}{m_{\rm tot}},
\end{align}
where $m_k$ is the mass in the $k^{th}$ bin and $m_{\rm tot}$ is the total mass. This process splits volumes in the dimension with the greatest amount of variation in the spacing between particles, effectively minimises the variation in particle density across any given cell volume. 

\par 
The cell size sets the background scale, below which we can robustly identify orbital clustering. We typically set $\Ncell=f_{\rm cell}N_{\rm H}$, where $f_{\rm cell}\sim0.01$ is the fraction of $N_{\rm H}$, the number of particles in the halo. This fraction is increased if $\Ncell\lesssim100$ up to a maximum of $\sim1/8N_{\rm H}$ in order to have an accurate dispersion tensor. 

\par
For each volume we calculate the centre-of-mass, centre-of-mass velocity and the velocity dispersion tensor:
\begin{align}
    \bar{\bf x}&=\frac{1}{M_{\rm cell}}\sum_k m_k{\bf x}_k,\label{eqn:meanpos}\\
    \bar{\bf v}&=\frac{1}{M_{\rm cell}}\sum_k m_k{\bf v}_k,\label{eqn:meanvel}\\
    \sigma^2_{i,j}&=\frac{1}{M_{\rm cell}}\sum_k m_k\left(v_{k,i}-\bar{v}_i\right)\left(v_{k,j}-\bar{v}_j\right)\label{eqn:meandispvel},
\end{align}
where $M_{\rm cell}$ is the mass contain in the cell and the sums are over all particles in the cell. The velocity quantities are interpolated to a particle's position with a inverse-distance interpolation scheme using the cell containing the particle and the six neighbouring cells (those that share faces with the cell of interest): 
\begin{align}
    u(\mathbf{x}) = \sum_{i = 0}^{N}{ \frac{ w_i(\mathbf{x}) u_i } { \sum_{j = 0}^{N}{ w_j(\mathbf{x}) } } },\label{eqn:cellinterp}
\end{align}
where $u$ is the quantity we wish to determine at a position $\bf x$ based on cells with center-of-mass positions $\bar{\bf x}_i$, and $w_i(\mathbf{x}) = |\mathbf{x}-\bar{\mathbf{x}}_i|^{-1}$. 

\par 
We then calculate the logarithmic ratio for each particle $k$,
\begin{align}
    \R_{k}=\ln\left[f_{\rm l}({\bf v}_k|{\bf x}_k)/f_{\rm bg}({\bf v}_k|{\bf x}_k)\right].
\end{align}
As both quantities have noise, this noise must be taken into account to determine if a particle is an outlier of the background distribution and belongs to a substructure. Based on tests using smooth, spherical halos with density profiles ranging from cored isothermal to a steep $r^{-1.5}(1+r/a_{o})^{-1.5}$ generated by {\sc galactICs} \cite[][]{galactics1995,galactics,widrow2008} the $\R$-distribution is characterised by Skew-Gaussian: 
\begin{align}
    f_{\rm SG}&(\R;\bar{\R},\sigma_\R,s,A)=
    A\Biggl\{\exp\left[-\frac{\left(\R-\bar\R\right)^2}{2s^2\sigma_\R^2}\right]\Theta(\bar\R-\R)\notag\\&\quad+\exp\left[-\frac{\left(\R-\bar\R\right)^2}{2\sigma_\R^2}\right]\Theta(\R-\bar\R)\Biggr\},\label{eqn:skewgaus}
\end{align}
where $s$ is a measure of the skew or asymmetry, and $\Theta(x)$ is the Heaviside function. The skew arises from the biased estimator of $f_{\rm l}({\bf v}_k|{\bf x}_k)$. We fit a Skew-Gaussian to the binned distribution in order to accurately measure the mean and dispersion and calculate the normalised ratio:
\begin{align}
    \ELL_k\equiv(\R_k-\bar{\R})/\sigma_{\R}. \label{eqn:ell}
\end{align}
A particle is considered a significant outlier if $\ELL>1$. 

\paragraph{Linking particles:}
\label{sec:velociraptor:substructures:linking}
The next stage uses a phase-space algorithm to link particles. Particles $i$ \& $j$ are linked iff
\begin{subequations}
\label{eqn:linkingcriteria}
\begin{gather}
    \ELL_i,\ELL_j\geq\ELL_{\rm th} \label{eqn:linkingcriteria:ellcrit}\\
    \frac{({\bf x}_i-{\bf x}_j)^2}{\left(\alpha_{\rm x,S}\ellx\right)^2}<1,\label{eqn:linkingcriteria:ellxcrit}\\
    1/\Vr\leq  v_i/v_j\leq \Vr,\label{eqn:linkingcriteria:vrcrit}\\
    \cos\Thetaop\leq \frac{{\bf v}_i\cdot{\bf v}_j}{v_i v_j}\label{eqn:linkingcriteria:thetaopcrit},
\end{gather}
\end{subequations}
where $V_r$ is the velocity ratio threshold, and $\cos\Thetaop$ is the threshold on the cosine of the angle between the velocities. 

\par 
The first criterion limits the linking to dynamically distinct particles. The second criterion is the standard FOF criterion with the linking length scaled by a factor $\alpha_{\rm x,S}$. The next two criteria ensure that the particles have similar velocities. The reason we do not use a simple 6DFOF, i.e., $({\bf v}_i-{\bf v}_j)^2/\ellv^2<1$, is that tidal streams may have large velocities and dispersions. Consequently, scaling an allowed velocity dispersion, $\ellv^2$ is non-trivial. In total, this FOF algorithm has $4$ parameters, $\ELLth$, $\alpha_{\rm S}$, $\Vr$ and $\cos\Thetaop$.

\par
As with all FOF algorithms, poor choice of linking parameters can produce spurious structures. A threshold of $\ELL_{\rm th}\approx0$ includes all particles whereas $\ELL_{\rm th}\gg1$ would ensure few contaminants. The speed ratio, $\Vr$, has two limiting cases: $\Vr\approx1$ is conservative; and $\Vr\gg1$ is relaxed. The related velocity parameter $\cos\Thetaop$ has limits of $\cos\Thetaop\approx1$ (conservative) and $\cos\Thetaop\approx-1$ (relaxed). This also applies to $\alpha_{\rm x,S}$, with $\alpha_{\rm x,S}<1$ ($\alpha_{\rm x,S}>1$) a conservative (relaxed) choice. Conservative choices would ensure high purity but possibly miss substructures, whereas more relaxed will recover more particles at the cost of a lower purity and the inclusion of spurious groups. 

\par
To alleviate the issue of either using conservative values and missing substructures or relaxed conditions that ensure maximum recovery but low purity, we also employ a two stage approach. First we use conservative values for the FOF parameters to find an initial set of candidate substructures. The FOF criteria are then relaxed to link previously untagged particles neighbouring currently tagged particles, thereby recovering the less dynamically distinct/more diffuse portions of substructures. The thresholds in \Eqref{eqn:linkingcriteria} are changed to $\ELLth\rightarrow\ELLth/\gamma_{\ELL}$, $\Vr\rightarrow\gamma_{\Vr}\Vr$,$\Thetaop\rightarrow\gamma_{\Thetaop}\Thetaop$, and linking lengths increased to $\gamma_{\rm x,S}\alpha_{\rm x,S}\ellx$, where the $\gamma$'s are order unity and $\geq1$. To recover extended tidal features, $\gamma_{\rm x,S}=1/\alpha_{\rm x,S}$, i.e., the linking length used to identify entire halos. 

\par
For guidance on the initial conservative parameters, we appeal to probabilistic or physical arguments. To minimise contamination, we start with $\ELLth\approx2.5$. The $\alpha_{\rm x,S}\ellx$ linking-length parameter can significantly influence the results and, in the form used, there is no specific value to appeal to without prior knowledge. We argue for $\alpha_{\rm x,S}\sim1/2$, picking out the densest regions of substructures. The speed ratio should be of order unity so values of $\sim2$ are reasonable. For the opening angle we typically use $\Thetaop=18^{\circ}$. These specific values are based on tuning done in \cite{elahi2011} to recover subhalos and tidal tails using idealised simulations, though similar values will yield similar results. 

\par 
Note that using conservative criteria can artificially split substructures and relaxing the criteria can join groups, in some circumstances artificially. Therefore, as substructures are grown and new links identified, substructures are only joined if the number of new connections exceeds $f_{\rm merge,th}N_{\rm p,o}$ for either substructure, where $N_{p,o}$ is the original size of the substructure. The default fraction threshold is  $f_{\rm merge,th}=0.25$, though values close to unity are reasonable.

\par 
The FOF algorithm without criterion \Eqref{eqn:linkingcriteria:ellcrit} and some tuning is itself able to recover the central densest regions of subhalos with moderate purity but {\em this criterion is critical to identify subhalos with high purity and robustly recover tidal debris}. 

\paragraph{Cleaning the catalogue:}
\label{sec:velociraptor:substructures:cleaning}
As with all halo finders, the catalogue must be cleaned of spurious groups and links. A group's average $\langle\ELL\rangle$ value is a natural measure of significance. Purely artificial groups resulting from linking unrelated particles that are outliers due to random fluctuations are likely to have  $\langle\ELL\rangle$ within Poisson noise of the expected $\bar\ELL$ calculated using the background distribution and the threshold $\ELL_{\rm th}$ imposed. Thus, we require a group composed of $N$ particles have
\begin{align}
    \langle\ELL\rangle\geq\bar\ELL(\ELL_{\rm th})\left(1+\beta_{\ELL}/\sqrt{N}\right).
    \label{eqn:ellsig}
\end{align}
Here $\beta_{\ELL}$ is the required significance level, typically $\beta_\ELL\approx1$ and 
\begin{align}
    \bar\ELL=\frac{\int\limits_{\ELLth}^{\infty}x{\rm e}^{-x^2/2}dx}{\int\limits_{\ELLth}^{\infty}{\rm e}^{-x^2/2}dx}=\frac{\sqrt{\frac{2}{\pi}}{\rm e}^{-\ELLth^2/2}}{1-{\rm erf}\left(\ELLth/\sqrt{2}\right)}.
\end{align}
Groups not satisfying this criterion have particles removed in order of smallest $\ELL$ value until \Eqref{eqn:ellsig} is satisfied.

\par 
Additionally, groups can be pruned by an unbinding process\footnote{The common terminology of ``unbinding'' is a bit misleading as discussed in \cite{knebe2013a}. The bound state is determined instantaneously, typically neglects the background potential by treating objects in isolation, and uses a somewhat arbitrary velocity reference frame. Loosely unbound particles at a given instant will not immediately leave their host but remain in similar orbits as their host, drifting away over a dynamical time.}, where by particles deemed too unbound are removed. We calculate the potential energy $W$ of particles using a tree algorithm with groups treated in isolation, that is neglecting the surrounding tidal field. The instantaneous kinetic energy $T$ is calculated relative to the group's centre-of-mass velocity reference frame\footnote{By default, the code uses shrinking spheres to determine the centre-of-mass and uses the inner most $10\%$ of particles to determine the centre-of-mass velocity. This can significantly differ from the bulk velocity of a halo as discussed in \cite{rockstar}. \textsc{VELOCIraptor} can be configured to use either a bulk velocity or a centre-of-mass velocity when determining the boundness of particles.}. 

\par 
In most halo finders, a strict binding energy is used, where particles with $T+W>0$ are removed and potentials and centre-of-mass velocity frames are recalculated with each removal. This strict unbinding process is only truly necessary for configuration-based finders such as {\sc subfind}, where initial particle assignment to subhalos can be quite poor. Due to the initial step of identifying dynamically distinct particles, \textsc{VELOCIraptor} does not suffer from this issue, allowing the binding criterion to be greatly relaxed in order to identify tidal debris. 

\par 
Therefore, to retain tidal debris if desired, we use a modified binding energy criterion, removing particles with 
\begin{align}
    \beta_{\rm E}T+W\geq0,\label{eqn:binding}
\end{align}
in order of least bound. For self-bound subhalos, $\beta_{\rm E}\approx0.95$ is ideal, retaining some loosely unbound particles that would not immediately drift away from their subhalo host\footnote{Consider particles orbiting inside an NFW potential representing the subhalo near the virial radius, where orbital time are $\gtrsim 1$Gyr. Particles with kinetic energies of $T=W/\beta_{\rm E}$ compared to $T=W$ for $\beta_{\rm E}=0.95$ will have apocentres that are $\lesssim10\%$ larger. These inflated radii are typically still within the tidal radius of a NFW subhalo orbiting inside a larger, less concentrated NFW halo, at least for orbital distances of $\gtrsim 0.5R_{200\rho_c,{\rm host}}$. Only once $\beta_{\rm E}\lesssim0.9$ do apocentres increase significantly by $\gtrsim50\%$, with apocentres likely outside the tidal radius.}. To retain tidal debris with high purity we find that $\beta_{\rm E}\gtrsim0.2$ works well \cite[based on tests presented in][]{elahi2013a}. One can also require that the group as a whole have some fraction of completely bound particles where $T+W\leq0$, $f_{\rm E}$.

\par
The total mass assigned to subhalos typically only changes by few percent for $0.95\lesssim\beta_{\rm E}\lesssim1$. This is well within the differences of $10-20\%$ observed between different (sub)halo finders \cite[][]{onions2012,knebe2013a}, which arise from subtle differences in the kinetic reference frame used and how potentials are calculated. We argue that unless one is interested in tidal debris, the binding criterion be set to $0.95\leq\beta_{\rm E}\leq1$, although one can always recover the formally self-bound mass in the output from the code for any $\beta_{\rm E}$.

\par 
Finally, groups must be composed of $N\geq N_{\rm min}$ particles. Typically we set $N_{\rm min}=20$.

\subsection{Core Search \& Major Mergers}
\label{sec:velociraptor:mergers}
Major mergers occur when two approximately equal mass objects (within a factor of a few) coalesce. These events present a uniquely difficult problem for many halo finders. Many configuration-space based finders will artificially shrink one of the objects, designating it a subhalo, while the other object will be artificially larger and be designated a host. The subhalo/halo designation and the mass can switch between objects. Phase-space based finders are in principle less prone to this swapping (see \citealp{behroozi2015a} for a discussion of major mergers; see \citealp{muldrew2011} for examples of the short-comings of configuration-space halo finders). 

\par 
During a major merger the ``halo'' consists of two (or more) overlapping distributions in phase-space containing similar amounts of mass. Our orbit clustering approach will not be able to disentangle the merging halos if the secondary halo is significantly larger than $f_{\rm cell}N_{\rm H}$ particles. In such an instance, the background will consist of the merging halo that we are trying to separate. 

\par
\begin{figure}
\centering
\begin{tikzpicture}[
    roundnode/.style={rectangle, rounded corners = 5pt, draw=ForestGreen, very thick, text width=0.35\textwidth,align=center},
    roundnode2/.style={rectangle, rounded corners = 5pt, draw=LightGreen, very thick, text width=0.3\textwidth,align=center},
    squarednode/.style={rectangle, draw=Cerulean, very thick, text width=0.4\textwidth, align=center},
    node distance=5mm
    ]
    \node[squarednode](searching){
    Identify field halos large enough to be searched for merger remnants using phase-space information \& search these $n_{\rm s}$ structures.
    };
    \node[roundnode](haloprop)[below=of searching]{
    Calculate $\bar{\bf X}(\Ncell),\Sigma_X(\Ncell)$ using all $n_{\rm bg}$ ``background'' particles of the (sub)halo.
    };
    \node[roundnode2](coresearch)[below=of haloprop]{
    Search ``background'' particles using 6DFOF for cores.
    };
    \node[roundnode2](coresearch2)[below=of coresearch]{
    Store largest group as new ``background'', if number of cores, $n_{\rm c}>1$ store secondary groups as candidate cores.
    };
    \node[roundnode2](coresearch3)[below=of coresearch2]{
    Update 6DFOF parameters $\ellx,\ellv$.
    };
    \node[rectangle,  rounded corners = 5pt, draw=LightGreen, ultra thick, rotate=90, yshift=15pt](loop2)[left=of coresearch2, anchor=north]{
    \begin{varwidth}{0.2\textwidth}
    While $n_{\rm c}>0,l<\Delta_{\rm C}$
    \end{varwidth}
    };
    
    \node[roundnode](coregrowthinit)[below=of coresearch3]{
    If any candidate cores were identified, start at deepest level where candidate cores were found. These are ``active'' cores and the particles belonging to the primary core at the previous level are ``active'' particles.
    };
    \node[roundnode2](coregrowth)[below=of coregrowthinit]{
    Calculate $\bar{\bf X},\Sigma_{{\bf X}}$ for ``active'' cores.
    };
    \node[roundnode2](coregrowth2)[below=of coregrowth]{
    Assign $n_{\rm a}$ ``active'' particles to closest core in phase-space.
    };
    \node[roundnode2](coregrowth3)[below=of coregrowth2]{
    Move to next level with ``active'' cores, add these to active list, adjust ``active'' particle list.
    };
    \node[rectangle,  rounded corners = 5pt, draw=LightGreen, ultra thick, rotate=90, yshift=15pt](loop3)[left=of coregrowth2, anchor=north]{
    \begin{varwidth}{0.2\textwidth}
    While $n_{\rm a}\neq0$ 
    \end{varwidth}
    };
    
    \node[roundnode](clean)[below=of coregrowth3]{
    Clean group catalogue of low significance cores and run unbinding routine.
    };
    \node[roundnode](newsearch)[below=of clean]{
    Identify substructures large enough to search for subsubstructures \& search these $n_{\rm s}$ structures.
    };
    \node[rectangle,  rounded corners = 5pt, draw=Green, ultra thick, rotate=-90, yshift=15pt](loop)[right=of coregrowthinit, anchor=north]{
    \begin{varwidth}{0.2\textwidth}
    Repeat if $n_{\rm s}>0$
    \end{varwidth}
    };

    \node[squarednode](result)[below=of newsearch]{
    Return list of substructures identified by search for phase-space dense cores and relation to field halos.
    };
    
    \draw[thick,->] (searching.south) -- (haloprop.north);
    \draw[->,thick] (haloprop.south) -- (coresearch.north);
    \draw[->,thick] (coresearch.south) -- (coresearch2.north);
    \draw[->,thick] (coresearch2.south) -- (coresearch3.north);
    \draw[->,thick] (coresearch3.south) -- (coregrowthinit.north);

    \draw[->,thick] (coregrowthinit.south) -- (coregrowth.north);
    \draw[->,thick] (coregrowth.south) -- (coregrowth2.north);
    \draw[->,thick] (coregrowth2.south) -- (coregrowth3.north);
    \draw[->,thick] (coregrowth3.south) -- (clean.north);

    \draw[->,thick] (clean.south) -- (newsearch.north);
    \draw[->,thick] (newsearch.south) -- (result.north);

    \draw[-,thick] (newsearch.east) .. controls +(right:7mm) .. (loop.east);
    \draw[<-,thick] (haloprop.east) .. controls +(right:7mm) .. (loop.west);
    \draw[<-,thick] (coresearch.west) .. controls +(left:7mm) .. (loop2.east);
    \draw[-,thick] (coresearch3.west) .. controls +(left:7mm) .. (loop2.west);
    \draw[<-,thick] (coregrowth.west) .. controls +(left:7mm) .. (loop3.east);
    \draw[-,thick] (coregrowth3.west) .. controls +(left:7mm) .. (loop3.west);

\end{tikzpicture}
    \caption{Activity chart for search for cores and identifying mergers.}
    \label{fig:merger:scheme}
\end{figure}
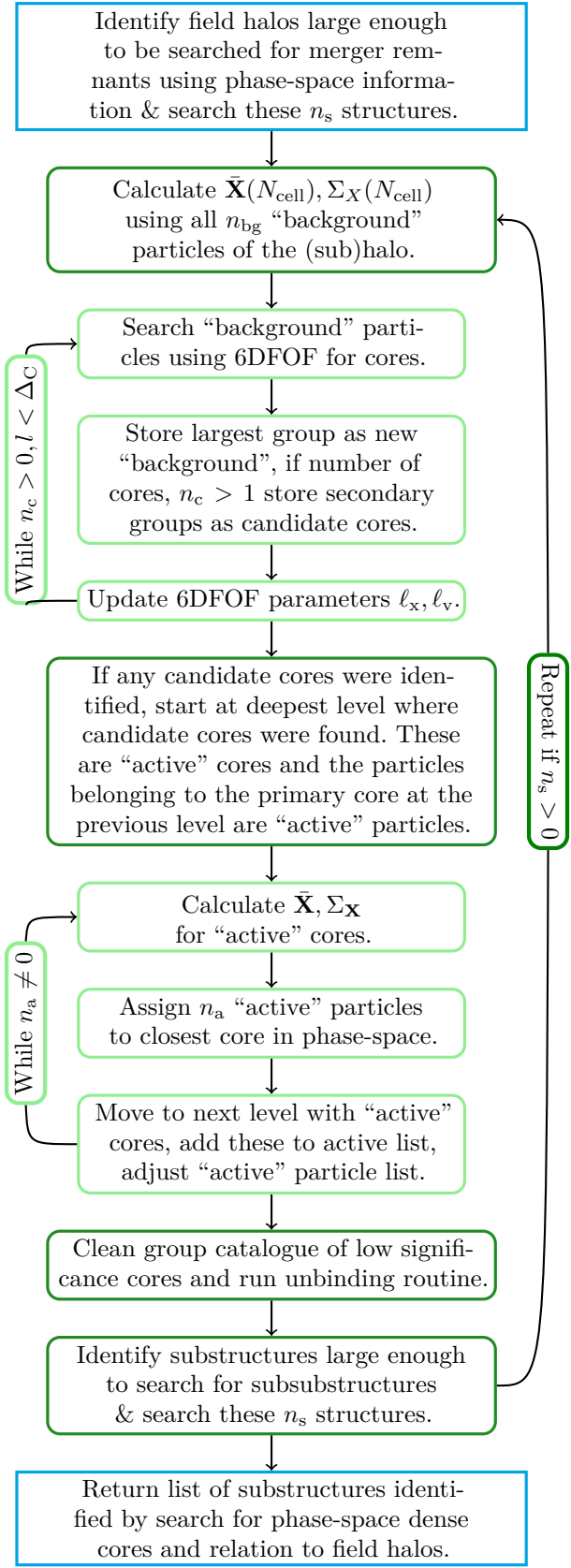

We disentangle mergers (both major and minor with mass ratios of $\gtrsim f_{\rm cell}$) by appealing to the properties of the dynamically cold, dense core of halos. An early version of this method was used in \cite{behroozi2015a}. Here we describe in full this new addition to \textsc{VELOCIraptor}. We search background particles not belonging to any substructure for these cores using a iterative, conservative 6DFOF and then proceed to grow them to reconstruct the mass as shown in \Figref{fig:merger:scheme}, taking inspiration from \textsc{rockstar} \cite[][]{rockstar}. 

\paragraph{Core Identification:}
\label{sec:velociraptor:mergers:cores}
We begin by searching the ``background'' particles of a halo, those not in substructure, using a conservative 6DFOF for groups larger than than some fraction $f_{\rm C}$ of $N_{\rm H}$ the number of particles in the halo. The linking lengths, $\ellx$ \& $\ellv$ here are based on the original halo linking length and the halo velocity dispersion respectively. This search is repeated with configuration- \& velocity-space linking lengths iteratively shrunk and the ``background'' particles list updated for each loop:
\begin{align}
    \ell_{\rm x,C}=\alpha_{\rm x,C}^l\ellx, \qquad
    \ell_{\rm v,C}=\alpha_{\rm v,C}^l\ellv, \label{eqn:coreloop}
\end{align}
where $\alpha_{\rm x,C},\alpha_{\rm v,C}<1$ and $l$ is the loop number. 

\par 
The ``background'' for each successive iteration is defined as the largest 6DFOF group identified in the previous iteration, the so-called ``primary core''. If at any point, more than a single group is identified, all but the largest are stored as candidate ``cores''. We loop until no groups are found (no background to search) up to a maximum desired number of iterations, $\Delta_{\rm C}$. The code can also alter the minimum number of particles a group must contain at a given iteration $l$ to $N_{\rm min,C}=\alpha_{\rm N,C}^l f_{\rm C}N_{\rm H}$.

\paragraph{Core Growth \& Mass Reconstruction:}
\label{sec:velociraptor:mergers:growth}
If more than a single ``core'' has been identified, the next step is to assign all untagged halo particles to these candidate ``cores'' and the ``primary core''. We start at the last iteration at which multiple groups were found, setting these ``cores'' and ``primary core'' as ``active''. Phase-space dispersion tensors are calculated for these active cores:
\begin{align}
    \bar{\bf X}&=\frac{1}{M_{\rm core}}\sum\limits^{N_{\rm core}}_k m_k {\bf X}_k,\label{eqn:coremeanphase}\\
    \sigma_{\bf X_{i,j}}&=\frac{1}{M_{\rm core}}\sum\limits^{N_{\rm core}}_k m_k\left(X_{k,i}-\bar{X}_i\right)\left(X_{k,j}-\bar{X}_j\right). \label{eqn:coredispphase}
\end{align}
We then assign untagged particles that were searched at this iteration, ``active background particles'', to the closest active core in phase-space. The distance used is:
\begin{align}
    D^2_{k,n}=({\bf X}_k-\bar{\bf X}_n)\Sigma^{-1}_{{\bf X},n}({\bf X}_k-\bar{\bf X}_n),
    \label{eqn:coredist}
\end{align}
where here we show the distance of particle $k$ to a core $n$ and $\Sigma$ is the matrix representation of $\sigma_{\bf X_{i,j}}$.

\par
Once all active particles at the current level are assigned, we then move up to the previous iteration and assign particles. If cores are present at this iteration, they are added to the active core list and we proceed as outlined above. We repeat the process till all particles not associated with substructure have been assigned to a core. 

\par 
This method is similar to assigning particles based on a Gaussian mixture model\footnote{A Gaussian mixture model is a probabilistic model that assumes data points are drawn from a mixture of a finite number of Gaussian distributions with unknown parameters. There are several techniques used to iteratively determine the number of Gaussians and their properties that describe the data using the Bayesian Evidence in some form. Data points can be assigned to the Gaussian with the highest probability of producing the data point, thereby classifying the data.}, but less time-consuming as we do not calculate full likelihoods. It also has the added advantage that we do not require each distribution to be characterised by a single global dispersion tensor. 

\par 
The use of phase-space tensor based distance also has an advantage over algorithms that use a simple 6DFOF-like distance metric (see \Eqref{eqn:fof6d}, e.g., {\sc rockstar}) as it does not impose a spherical distribution, nor ignore covariance between positions and velocities. That is not to say that for moderately aspherical distributions typical of halos, using scalar dispersions performs poorly but that results can be improved using dispersion tensors. 

\par 
We compared assigning particles using dispersion tensors to dispersion scalar using simple models composed of overlapping multivariate Gaussians. We draw particles from several $n$-dimensional multivariate Gaussian distributions with means roughly separated by $\sim1-3\sigma$ from each other, and with each sub-population containing similar numbers of members. Initial dispersion scalars and tensors are determine using 100 particles and then assign particle group membership using the relevant distance in single step. We find tensor-based distance assignment results in groups of higher purity, that is a higher fraction of correctly identified members. There is also a reduction in the group-to-group scatter in purity. The amount of improvement depends on the asphericity of the distributions, with increase of a few percent or more. More aspherical distributions have larger increases in purity as well as the fraction of the group recovered. Iterating this process improves the results. 

\par 
For example, consider particles drawn from two Gaussian distributions, one spherical, the other quite aspherical (with minor axis ratio of $0.03$), separated by a phase-space tensor normalised distance of $\sim2$. Assignment using the dispersion scalar distances results in a purity of $0.76$ \& $0.92$ for the spherical and aspherical populations respectively. Using tensor based distances improves the purity to $0.79$ \& $0.93$ respectively. The recover fractions are similarly improved from $0.94$ \& $0.70$ to $0.95$ \& $0.76$ respectively. 

\paragraph{Cleaning the Catalogue:}
\label{sec:velociraptor:mergers:cleaning}
We clean the candidate core list of spurious objects prior to core growth by requiring that the distance of a core $n$ to the primary core $p$ identified at the same point to be significant, 
\begin{align}
    D^2_{p,n}\geq\beta_{\rm C},
    \label{eqn:coresig}
\end{align}
where the distance is based on the secondary core's phase-space tensor using \Eqref{eqn:coredist}, and $\beta_{\rm C}$ is the significance. The substructures after core growth are then processed by the unbinding procedure (see \Eqref{eqn:binding}).

\subsection{Substructure and Baryons}
\label{sec:velociraptor:baryons}
Assigning baryonic particles to substructure or identifying baryonic substructures depends on the mode of operation. We discuss the two principal modes here. 
\paragraph{Substructure in DM+Baryons mode:}
In this mode, baryons have already been assigned to a FOF envelop. For each FOF envelop, baryons are assigned to the group of the DM particle that is closest in phase-space using a simple phase-space metric
\begin{align}
    D^2_{{\rm B,DM}}=({\bf x}_{\rm B}-\bar{\bf x}_{\rm DM})/\ellx+({\bf v}_{\rm B}-\bar{\bf v}_{\rm DM})/\sigma_v, \label{eqn:baryondist}
\end{align}
where $\sigma_v$ is the typical velocity dispersion of structures found\footnote{A more complex phase-space metric could be used, where the dispersion depends on the FOF halo being searched or even a full tensor but the extra computational cost does not drastically improve the initial particle assignment. This is particularly true when the initial baryonic assignment is processed by an unbinding routine.}. 

\paragraph{Substructure in Galaxies+Baryons mode:}
The process used to identify DM substructures is ill suited to separating interacting galaxies as stars are constantly being formed and there need not have a well defined background. Instead interacting galaxies are separated using the core search as outlined in \Secref{sec:velociraptor:mergers} \cite[see][for details]{canas2018a}. Once interacting galaxies have been separated, the same assignment scheme is used as in the \textit{DM+Baryons} mode to assign other baryonic particles (gas and black hole particles) to the nearest star particle. 

\subsection{Halo Properties}
\label{sec:velociraptor:properties}
The code calculates a large number of bulk properties for each object (see \Tableref{tab:velociraptor:properties} for an almost complete list; the exact number of properties calculated depending on input). Calculating properties is complicated by the presence of substructure. Should substructures be excluded or included? The answer depends on the scientific goal. For following the evolution of objects across cosmic time using halo merger trees for input into SAMs, ideal masses are likely that of particles belonging exclusively to the object, whether halo or subhalo. This avoids abrupt changes in mass as an object transitions from a halo to a subhalo. For lensing, one is likely interested in the total mass within some region. 

\par 
\textsc{VELOCIraptor} allows some flexibility: masses can either be calculated using particles exclusive to the object, or, for halos one can include substructures. Inclusive halo masses, such as commonly used spherical overdensity halo masses\footnote{An example would be $M_{200\rho_c}=4\pi\Delta\rho_c R_{\Delta\rho_c}/3$, where $\rho_c$ is the critical density, and $R_{\Delta\rho_c}$ is the radius enclosing an average density of $\Delta\rho_c$, where $\Delta=200$} can include particles belonging to substructures, the background and even neighbouring halos. Subhalos have exclusive masses, that is calculated using only particles belonging to the subhalo. Angular momentum, like mass can be calculated in a variety of was for halos. Other properties, such the maximum circular velocity, are by default are calculated using particles exclusively belonging to the object.

\par 
Another complication in bulk properties has to do with the phase-space position of a halo. The overall bulk motion of particles within the FOF envelop maybe offset from the motion of the central most bound regions particularly the motions of particles near the edge of the FOF envelop \cite[][]{rockstar}. By default, centre-of-mass positions are calculated using shrinking spheres till the the last sphere encloses $\sim10\%$ of the group's particles and velocities are calculated using this inner most $10\%$. These positions better characterise the orbital motion of halos, though it does not represent the overall bulk motion of mass. 

\par 
\textsc{VELOCIraptor} also outputs all the particle IDs in each structure so users can post-process data to calculate desired properties.

\subsection{Code Structure}
\label{sec:velociraptor:summary}
\textsc{VELOCIraptor} is a {\sc c++} code that uses OpenMP+MPI APIs for parallelisation but can be compiled in serial mode, solely with OpenMP, or solely with MPI. The code requires a configuration file (example are provided with the repository), input data and an output file name. 

\par 
The code has been designed to take the following types of N-body/Hydrodynamical input: {\sc HDF5}\footnote{Library can be found at \href{https://www.hdfgroup.org/}{https://www.hdfgroup.org/}.}; {\sc gadget} binary format \cite[][]{springel2005}; {\sc ramses} binary format \cite[][]{ramses}; and {\sc tipsy} binary format \cite[][]{tipsy}. For all input save {\sc tipsy}, \textsc{VELOCIraptor} extracts cosmological information and the spatial bounds for the particles. This information can be provided via the configuration file if not present in the input data. 

\par 
The spatial extent of the particle distribution must be provided for MPI domain decomposition, even for non-periodic input. This information can be provided either via the input data itself or via the configuration file. Currently implemented MPI domain decomposition scheme is a Binary Tree like splitting\footnote{A graph-partitioning scheme using the \textsc{metis} library \href{http://glaros.dtc.umn.edu/gkhome/metis/metis/overview}{http://glaros.dtc.umn.edu/gkhome/metis/metis/overview} is in the works}. 

\par 
It produces the following types of output formats: ASCII; custom binary format; {\sc HDF5} ({\em preferred}); and ADIOS\footnote{Library can be found at \href{https://www.olcf.ornl.gov/center-projects/adios/}{https://www.olcf.ornl.gov/center-projects/adios/}.} (alpha). The output files consist of two types: a collection of bulk properties for each group; and a list of the IDs of all particles belonging to each group. It can also produce a list of particle types and even information on the file and index each particle is located at, allowing for quick extraction of particle data for further follow-up analysis. We outline a sample of the bulk properties calculated in the appendix, \Tableref{tab:velociraptor:properties}.

\par 
There are a variety of configuration options available. We list the critical parameters in \Tableref{tab:velociraptor:params}, providing a more complete list and the specific code parameter key words in the appendix, \Tableref{tab:velociraptor:config}. We note that for most users, these default parameters will produce standard halo catalogs and subhalo need no alteration. Most users will simply alter the minimum number of particles per halo. For identifying tidal debris, the key parameter to change is the unbinding parameter, $\beta_{\rm E}$, which can be set to values of $\sim0.1-0.5$. We highlight parameters that are likely to be changed depending on the input simulations and the desired scientific outcome. 
\begin{table*}
\setlength\tabcolsep{2pt}
\centering\footnotesize
\caption{Key \textsc{VELOCIraptor} parameters}
\label{tab:velociraptor:params}
\begin{tabular}{@{\extracolsep{\fill}}l|l|c|p{0.65\textwidth}}
\hline
\hline
    & Name & Default Value & Comments\\
    \hline
    Field Halos & & & Related to field halo search. \\
    \hline
    & $\ellx$ & 0.2 
        & Base 3DFOF configuration-space linking length in units of inter-particle spacing. See \Eqref{eqn:fof3d}, \Eqref{eqn:fof6d} \footnote{For historical reasons, the code actually uses the substructure linking length to define the halo linking length, i.e., the code actually takes $\alpha_{\rm x, H}$ and $\ell_{\rm x, S}$ as input}. The typical value of 0.2 is the commonly used one for finding 3DFOF halos and corresponds to identifying overdensities of $\gtrsim100\rho_m$. \\
    & $\alpha_{\rm v}$ & 1.25 
        & Scaling of base 6DFOF velocity-space linking length $\ellv$ in units of local halo velocity dispersion. See \Eqref{eqn:fof6d}. For virialised objects, typical values should be of order unity. Using large values will effectively transform 6DFOF halos to 3DFOF halos. \\
    \hline
    Substructure & & & Related to velocity outlier substructure search. \\
    \hline
    & $\Nv$ & 32 
        & Number of particles used to estimate local velocity density.\\
    & $\Nse$ & 256
        & Number of neighbouring particles used to estimate local velocity density.\\
    & $f_{\rm cell}$ & 0.01 
        & Fraction of halo contain in cell used to estimate background velocity density. See \Eqref{eqn:fvbg}, Eq. (\ref{eqn:meanpos}-\ref{eqn:meandispvel}). \\
    & $\ELLth$ & 2.5 
        & Outlier threshold when linking particles. See \Eqref{eqn:linkingcriteria:ellcrit}.\\
    & $\alpha_{\rm x,S}$ & 0.5 
        & Scaling of the $\ellx$, halo linking length. See \Eqref{eqn:linkingcriteria:ellxcrit}.\\
    & $\Vr$ & 2 
        & Velocity ratio. See \Eqref{eqn:linkingcriteria:vrcrit}.\\
    & $\Thetaop$ & 0.1 
        & Opening angle in units of $\pi$. See \Eqref{eqn:linkingcriteria:thetaopcrit}.\\
    \hline
    Core Search \& Major Mergers & & & Related to core/major merger search.\\
    \hline
    & $\alpha_{\rm x,C}$ & 0.8  
        & Scaling of 6DFOF configuration-space linking length in core search. See \Eqref{eqn:coreloop}.\\    
    & $\alpha_{\rm v,C}$ & 1.0
        & Scaling of 6DFOF velocity-space linking length in core search. See \Eqref{eqn:coreloop}.\\
    & $n_{\rm loop,C}$ & 5
        & Number of loops to search for cores. \\
    & $f_{\rm C}$ & 0.02
        & Fraction of halo a core must contain. \\
    & $\alpha_{\rm N,C}$ & 1.2 
        & Scaling of $N_{\rm min,C}$ in core search. \\
    \hline
    Unbinding/Cleaning & & & Related to cleaning of structure catalogue and particles associated to structures. \\
    \hline
    & $N_{\rm min}$ & 20 
        & {\bf Key Parameter}: minimum number of particles an group must contain.  This parameter is likely the most often altered one depending on the science goal in question. For instance, for Smooth Particle Hydrodynamical simulations, objects composed of fewer particles than the number used to calculate the hydrodynamics are dominated by numerical artefacts and can be ignored. . \\
    & $\beta_{\rm E}$ & 0.95 
        & {\bf Key Parameter}: fraction of kinetic energy that potential energy must be for particle to be considered bound. See \Eqref{eqn:binding}. As the code will naturally recover the continuum of substructures, from well bound, physically overdense, dynamically cold subhalos to unbound, underdense, dynamically cold streams, this parameter needs to be set with the desired catalogue in mind. If the use desires a catalogue of bound subhalos, then this default value is acceptable. To include more tidal debris objects, this parameter should be decreased. \\
    & $\beta_\ELL$ & 1 
        & Significance of substructure's outlier average relative to noise. See \Eqref{eqn:ellsig}.\\
    & $\beta_{\rm C}$ & 1 
        & Significance of core relative to noise. See \Eqref{eqn:coresig}.\\
\end{tabular}
\end{table*}

\section{Results}
\label{sec:results:velociraptor}
Here we present how well halos/galaxies and substructures are identified. As input we primarily use a small cosmological N-Body simulation consisting of $512^3$ particles \cite[from the SURFS suite][]{elahi2018a}. The simulation details are presented in \Tableref{tab:sims}.
\begin{table}
\setlength\tabcolsep{2pt}
\centering\footnotesize
\caption{Simulation parameters}
\begin{tabular}{@{\extracolsep{\fill}}l|cccc}
\hline
\hline
    Name & Box size & Number of & Particle Mass & Softening \\
    & & Particles & &Length\\
    & $L_{\rm box}$ [$\Mpch$] & $N_p$ & $m_p$ [$\Msunh$] &  $\epsilon$ [$\kpch$] \\
\hline
    L40N512     & $40$  & $512^3$   & $4.13\times10^7$ & 2.6 \\
    L210N1536     & $210$  & $1536^3$ & $2.21\times10^8$ & 4.5 \\
\hline
\end{tabular}
\label{tab:sims}
\end{table}
For this analysis, we also make use a halo merger tree builder, \textsc{TreeFrog} \cite[][]{treefrogpaper}. This related software is a so-called ``Tree Builder', software that takes as input halo catalogs across cosmic time and reconstructs the history of a halo, producing halo merger trees. Details of how \textsc{TreeFrog} reconstructs halo merger trees can be found in \cite{treefrogpaper}. Here we summarise the salient points: the code uses particle IDS and the group to which they belong to compare one snapshot to the next, identifying descendants by maximising a merit function that effectively links halos at a time $t_1$ to halos found a later time that share the largest number of most bound particles. We also compare results to three different (sub)halo finders: \textsc{ahf} \cite[]{ahf}, a configuration-space based finder; \textsc{rockstar} \cite[]{rockstar}, a phase-space finder; and \textsc{hbt+} \cite[]{han2018a} a 3DFOF tracker that uses 3DFOF halos found across all snapshots and tracks particles assigned to 3DFOF halos as they enter larger 3DFOF halos to build a halo merger tree as well as a subhalo hierarchy. 

\par 
We start by looking at the identification and decomposition of individual objects and then look at the statistical properties of the ensemble population extracted from our simulations. We use a particle limit of $N_{\rm min=20}$ and focus on self-bound objects, that is we use $\beta_{\rm E}=0.95$ (see \Eqref{eqn:binding}).

\subsection{Individual Halo}
\label{sec:results:velociraptor:halo}
\begin{figure*}
    \includegraphics[width=0.98\textwidth,trim=0.cm 0.cm 0cm 0cm, clip=true]{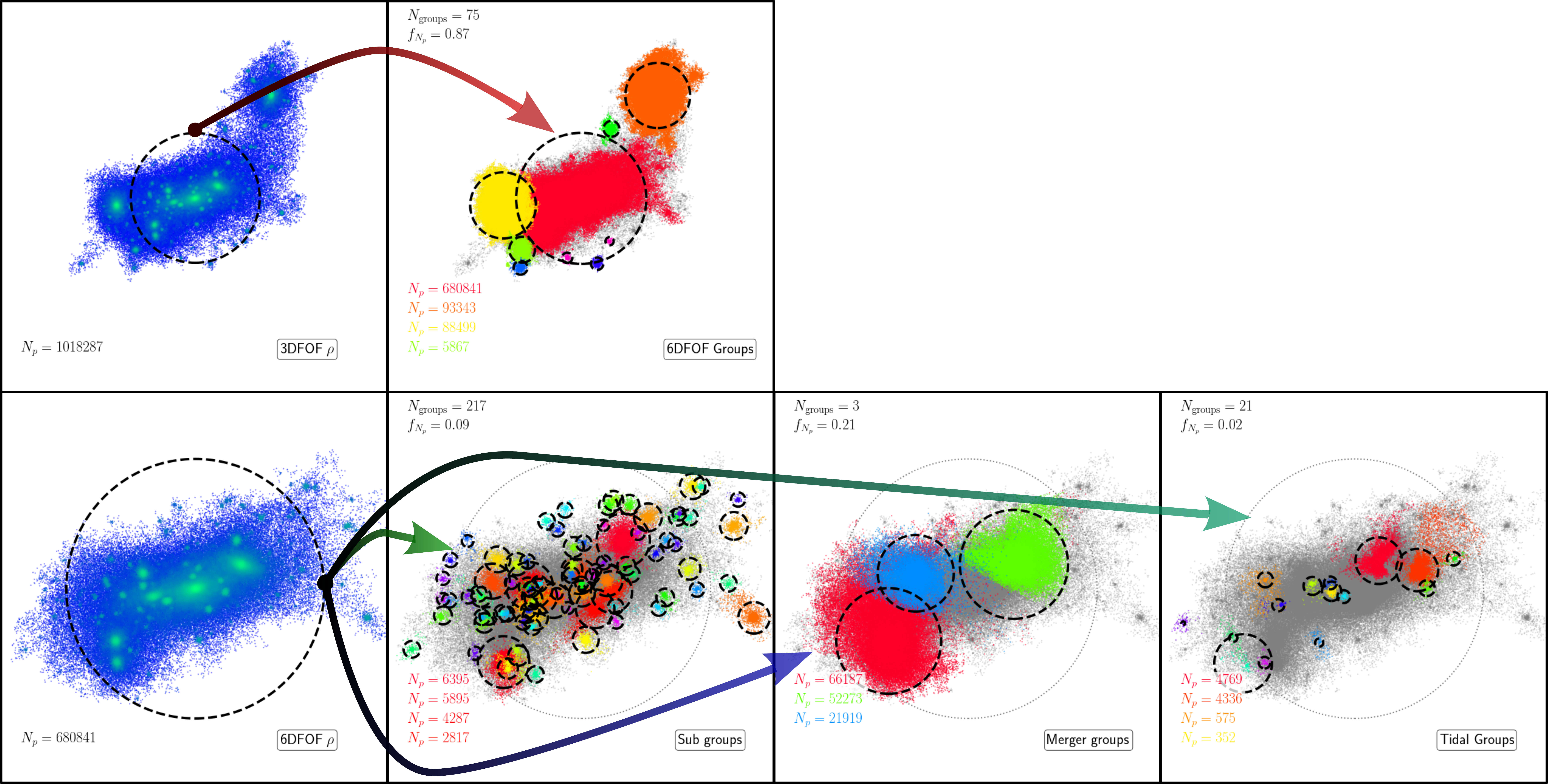}
    \caption{{\bf Halo Decomposition}: We show the process of running the routines that decompose an initial FOF candidate into 6DFOF Halos (top row), followed by the search for substructure (using \Secref{sec:velociraptor:substructures}) and major mergers (using \Secref{sec:velociraptor:mergers}) in the largest 6DFOF halo (bottom row, red 6DFOF halo seen in top right panel). The bottom panel shows the application of substructure finding (green arrow), core identification and grow for mergers (purple arrow), and the substructures identified when the self-boundness criteria are relaxed to find tidal debris (teal arrow). For each object we show $R_{\Delta\rho_c}$ by a dashed black circle. In the left column, particles are colour-coded according to the 3D density going from blue to green in increasing density. In the other panels (group sub-panels), particles are colour-coded by the group to which they belong. In these group sub-panels: we limit the number of groups displayed to those composed of more than 100 particles for clarity; list the total number of groups; the fraction of mass in these groups; the number of particles for the 4 largest such groups; and show the parent halo's particles and $R_{\Delta\rho_c}$ with gray points and a gray circle respectively.}
    \label{fig:halodecomposition}
\end{figure*}
Figure \ref{fig:halodecomposition} shows a 3DFOF halo extracted from the L40N512 simulation and how each step in VELOCIraptor decomposes the candidate/parent object. In this example, we use a large halo composed of $\approx10^{6}$ particles identified at $z=0$ with a 3DFOF mass of $4.2\times10^{14}\Msunh$ and a mass $M_{\Delta\rho_c}=2.7\times10^{14}\Msunh$, where $M_{\Delta\rho_c}=4\pi\Delta\rho_c R_{\Delta\rho_c}/3$, $\rho_c$ is the critical density, and $R_{\Delta\rho_c}$ is the radius enclosing an average density of $\Delta\rho_c$, where $\Delta=200$, commonly referred to as the virial mass. This 3DFOF object was identified using the standard linking length of $\ell_x=0.2 L_{\rm box}/N_p$, where $L_{\rm box}/N_p$ is the inter-particle spacing. 

\par
The initial 3DFOF halo clearly consists of several large density peaks, some of which are well outside the virial radius centred on the largest density peak. All the density peaks would be considered subhalos of the FOF envelop, save for the peak that has the largest mass associated with it, which is considered the parent halo. This subhalo/halo definition is less than ideal as some of the larger peaks are well outside the virial radius. Moreover, some of these peaks are spuriously linked to the primary via a thin particle bridge by the FOF algorithm. This example illustrates the need for more sophisticated algorithms. 

\par 
Applying the 6DFOF algorithm separates the initial 3DFOF candidate into 75 (bound) groups, 3 of which are composed of $\gtrsim10\%$ of the original 3DFOF object's particles. Approximately $87\%$ of the original 3DFOF object's particles are still within a group, with the largest object containing $68\%$ and having approximately the same virial mass as the original 3DFOF. The 6DFOF algorithm produces a better mapping of a FOF object to the physical definition of a halo, that of a virialised overdensity. 

\par 
The largest 6DFOF halo itself appears to contain at least 4 large density peaks and numerous smaller ones. If we search for substructure by identifying locally dynamically distinct particles and linking them with a phase-space algorithm (method outlined in \Secref{sec:velociraptor:substructures}) we find 217 groups, containing $\approx9\%$ of the mass of the halo, the few largest of which each contain $\approx1\%$ of the total halo's mass. 

\par
The largest density peaks within the 6DFOF are separated into 3 groups plus the main halo by the core search (see \Secref{sec:velociraptor:mergers}). These objects, remnants of minor/major mergers, contains $21\%$ of the initial host halo's mass, with the smallest containing $3\%$ and the largest $9\%$. The second largest merger remnant happens to be close to the main halo, making particle assignment during the core growth phase non-trivial, particularly for the outer regions that overlap in phase-space with the host. The sharp boundary between the object and the main halo is a result of a compromise between computational efficiency and rigour as we use few steps to grow cores and a global phase-space tensor to assign particles based on their distance to the cores's centre-of-masses. Finer steps would reduce the sharpness of this transition but as it effects small amounts of mass, extra steps are unnecessary. 

\par 
For comparison, other methods find similar amounts of mass, though there are some differences. \textsc{hbt+}, which tracks halos, assigns the least amount of mass to the most distant object. \textsc{ahf} underestimates the mass of the most central object relative to all other finders, expected given its configuration-space approach. \textsc{rockstar}, which has a similar approach to that outlined in \Secref{sec:velociraptor:mergers}, returns similar, if typically larger masses. Both \textsc{VELOCIraptor} and \textsc{rockstar} also give similar results to a full Gaussian mixture model using centre-of-mass of the cores as initial guess\footnote{We use an implementation in \textsc{SciKit} \textsc{Python} package that uses variational inference which maximizes a lower bound on model evidence (including priors) instead of data likelihood.}. 

\par
\begin{figure}
    \centering
    \includegraphics[width=0.49\textwidth,trim=0.75cm 1.cm 1cm 0.75cm, clip=true]{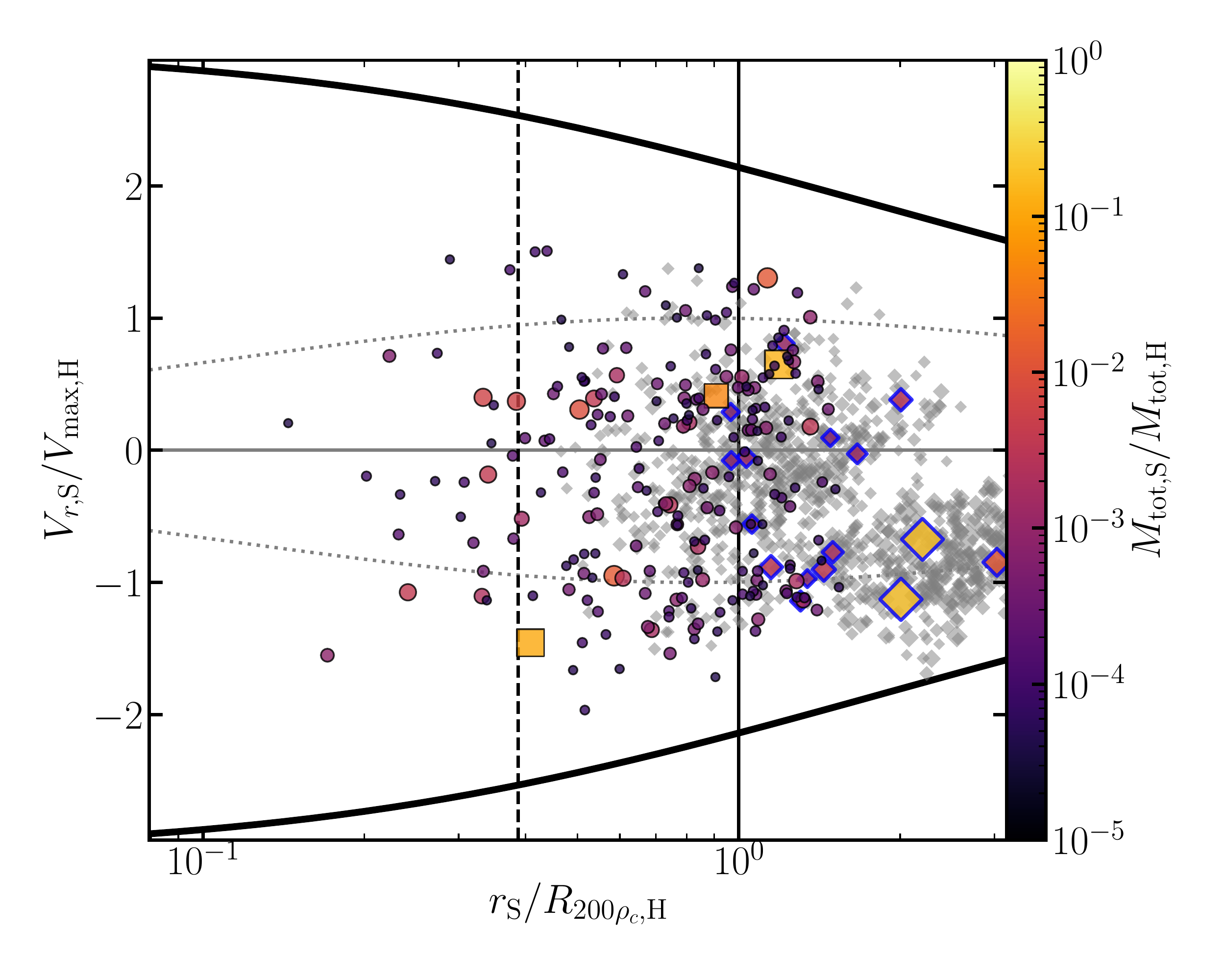}
    \caption{{\bf Phase-space distribution of substructures in the halo:} We plot the radial position and velocity (scaled by the host halo properties) of all substructures found in the example 6DFOF halo with points colour-coded by mass (and scaled by mass as well). We plot minor/major mergers as square points and all other substructures as circles. We also plot the escape velocity envelop (solid black lines), circular velocity envelop (dotted gray lines) and the scale radius of the NFW concentration (vertical dashed line). We plot the large 6DFOF halos that were part of the initial 3DFOF envelop as diamonds with blue outlines, with points color coded and scaled by mass. Finally we also plot any objects not considered part of the initial 3DFOF and within $3R_{200\rho_c}$ as gray diamonds to show the halo population (and subhalos in other halos) in the surrounding environment.}
    \label{fig:halosubhalophaseexample}
\end{figure}

The phase-space distribution of these objects within of the parent halo is presented in \Figref{fig:halosubhalophaseexample}. Here we focus on the objects found within the 6DFOF envelop and use the total mass exclusively assigned to an object, $M_{\rm tot}$\footnote{\textsc{VELOCIraptor} does calculate overdensity masses such as $M_{200\rho_c}$ for subhalos. However, these masses are calculated treating the object in isolation unlike the calculation for field halos as using all particles within a spherical region is not as physically meaningful for an object that itself resides in an overdensity.}. The relative velocities and radial distances of the subhalos are scaled by maximum circular velocity of the host and its virial radius. We also show the largest halos that were separated by the 6DFOF from the initial 3DFOF envelop. 


\par 
The radial motions (as well as the total relative velocity) of all subhalos belonging to the 6DFOF envelop are well within the escape velocity envelop. For this particular halo, the parent 3DFOF halo would have linked together several objects that are on first infall and lie outside the virial radius, again pointing to a better mapping between a 6DFOF object and a virialised overdensity, a.k.a, a halo. For example, the typical apocentre for particles orbiting a halo is $\sim1.6-1.9R_{200{\rm crit}}$ \cite[though the exact value depends on the mass accretion rate of a halo and the rarity of the halo, for $75\%$ of a halo's particles apocentres are $\approx 1.0-1.2R_{200\rho_m}$, where $R_{200\rho_m}\approx1.6R_{200\rho_c}$, see][]{diemer2017b}. The two largest objects separated by the 6DFOF algorithm are well outside the virial radius at similar distances of $\approx 2R_{200\rho{c,{\rm H}}}$. However, they have large infalling radial velocities of $\approx-0.9V_{\rm max,H}$, significantly different from most particles that have completed at least one orbit of their host halo. Following their history using a halo merger tree \cite[built with \textsc{TreeFrog}][]{treefrogpaper}, we see that they are on first infall, as are most of the halos within the surrounding environment\footnote{The reason for the large number of background gray points is that there are a large number of loosely bound, poorly resolved 6DFOF halos around the main 6DFOF halo and the three infalling halos are quite rich, containing lots of substructure.} (as seen by the gray diamonds with negative velocities in \Figref{fig:halosubhalophaseexample}). 

\par
\begin{figure}
    \centering
    \includegraphics[width=0.4\textwidth,trim=0.cm 0.cm 0cm 0cm, clip=true]{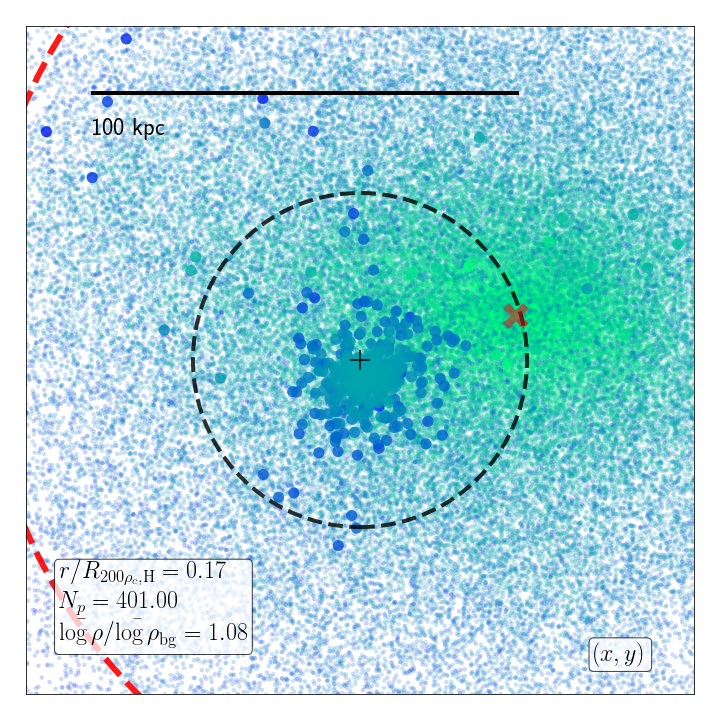}
    \includegraphics[width=0.4\textwidth,trim=0.cm 0.cm 0cm 0cm, clip=true]{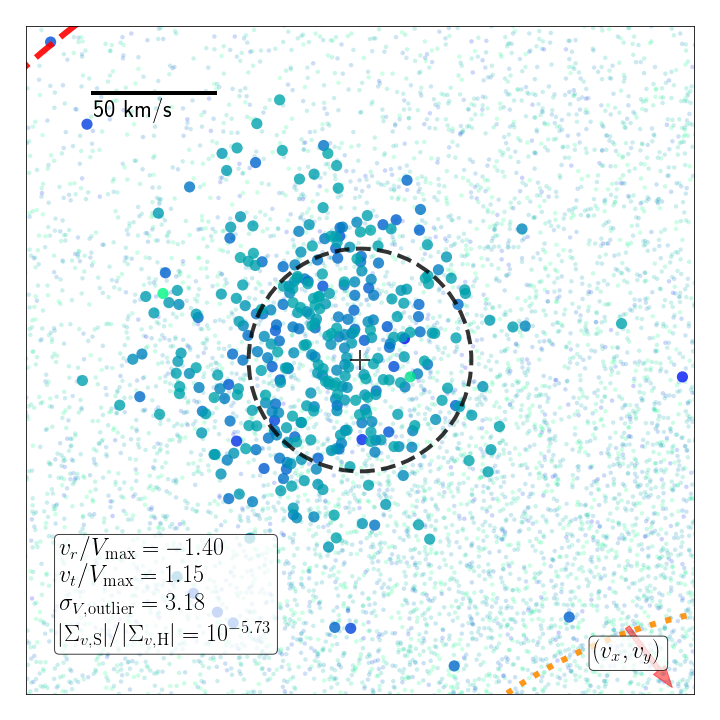}
    \caption{{\bf Inner subhalo:} We show a subhalo identified within the scale radius of a host halo. We plot its configuration space (top) and velocity space (bottom) distribution. Particles belonging to the subhalo are plotted as large circles, the background halo as small points, with points colour-coded by $\log\rho$, increasing in density going from blue to green. In the top panel, we mark the centre-of-mass by a '+', its $R_{200\rho_c}$ by a dashed circle. We also mark the center of the parent halo by a 'x' and also show the scale radius by a dashed red circle (seen in the left corners). In the bottom panel, we plot the centre-of-mass velocity with a '+' and $V_{\rm max}$ by a dashed circle. The parent halo's centre-of-mass velocity is off the plot in the direction of the red arrow. We also plot the parent halo's $V_{\rm max,H}$ by a red dashed circle (seen in the top corner) and also plot an ellipse centred on the mean velocity of the background particles in the nearby volume with its size scaled by the standard deviation (seen in lower-right corner). For both panels we plot a ruler to give a sense of scale.}
    \label{fig:innersubhaloexample}
\end{figure}

 
The inner most subhalos highlight the benefit of a phase-space finder. As an example, we focus on the large infalling subhalo located at $\approx0.2R_{200\rho_c,{\rm H}}$ and its surroundings, presented in \Figref{fig:innersubhaloexample}. In configuration space, the subhalo has a similar density to the background halo. It is only in velocity space that the subhalo becomes readily apparent. The object is a local velocity outlier as it lies outside the local velocity dispersion. The extent to which this object centre-of-mass motion $V_{\rm S}$ relative to the local surroundings is an outlier is given by
\begin{align}
    \sigma_{V,{\rm outlier}}\equiv
    \left[
    \left({\bf V}_{\rm S}-\bar{\bf v}_{\rm bg}\right)
    \Sigma_{v,{\rm bg}}^{-1}
    \left({\bf V}_{\rm S}-\bar{\bf v}_{\rm bg}\right)
    \right]^{1/2}
\end{align}
where $\bar{\bf v}_{\rm bg}$ and $\Sigma_{v,{\rm bg}}$ are the local mean velocity and velocity dispersion tensor. We find that its centre-of-mass velocity is a $\gtrsim3\sigma$ outlier of the local velocity distribution. Moreover, the particles belong to the object are far more clustered about its velocity than the expectation, with the ratios of the dispersion tensors giving $2\times10^{-6}$. 

\par 
To compare the particles belonging to the substructure to the background, we randomly sample the background particles 1000 times using the same number of particles belonging to this subhalo in a region centred on the subhalo within a radius of $1.5R_{\rm 200\rho_c}$. We find velocity differences of $\sigma_{V,{\rm outlier}}=3.27\pm0.18$, dispersion ratios of $|\sigma_{\rm S}|/|\Sigma_{\rm bg}|=(1.6\pm0.2)\times10^{-6}$ and density ratios of $\left\langle{\log\rho_{\rm S}}\right\rangle/\left\langle{\log\rho_{\rm bg}}\right\rangle=1.02\pm0.02$. The object's mean density is similar to the background yet the subhalo's centre-of-mass velocity is a significant outlier and its velocity dispersion is much colder. 

\par 
The low density contrast does not necessarily mean that this object cannot be recovered by configuration-space finders. For instance, \textsc{AHF} does recover this object, however, the object proceeds to shrink as it enters deep within the host. Moreover, the initial collection of particles within the density peak will contain both particles belonging to the subhalo and those of the background, which must be carefully pruned by an unbinding process. By using velocity information, the particles belonging to the object can be robustly separated from the background, particularly the more underdense outer regions, minimising the amount of cleaning that must be done.

\par
\begin{figure}
    \centering
    \includegraphics[width=0.45\textwidth,trim=1.cm 4.cm 1.25cm 5.5cm, clip=true]{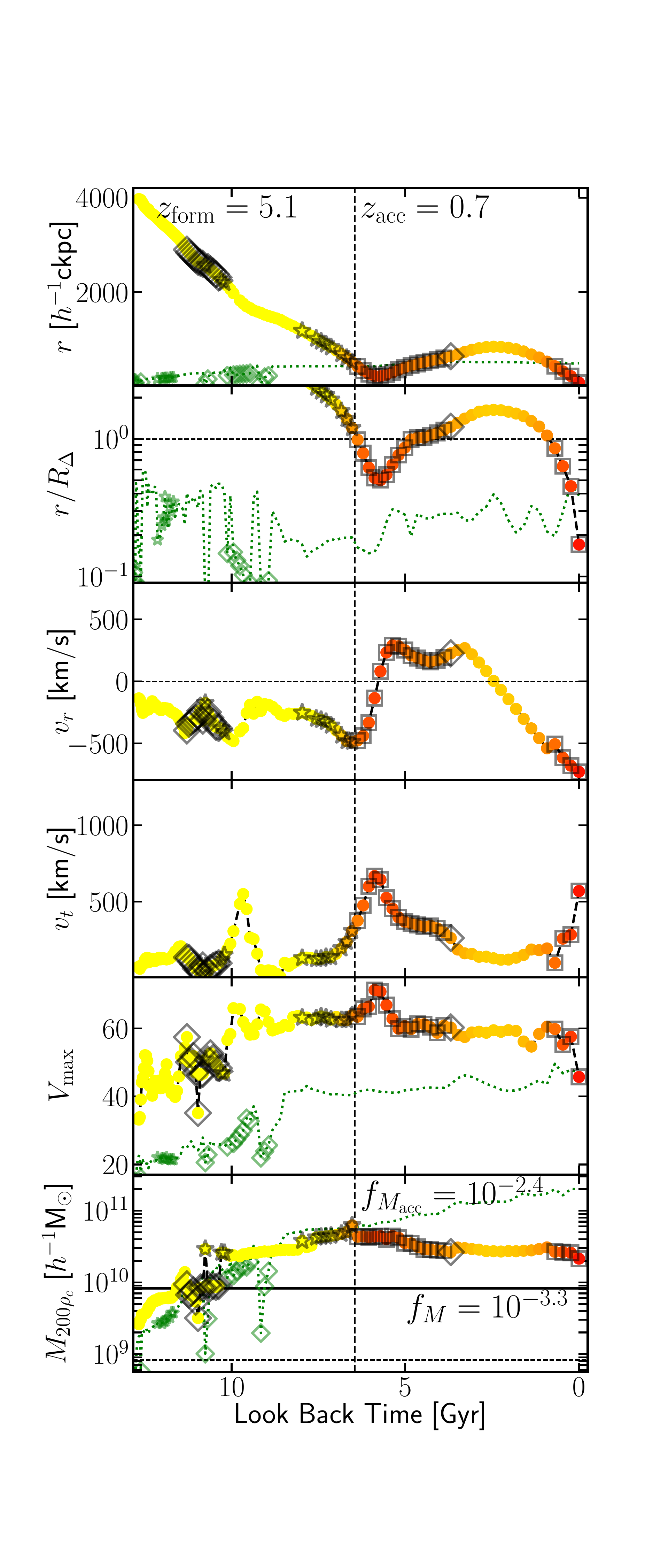}
    \caption{{\bf Reconstructed Subhalo Orbital \& Evolution:} We plot the orbit \& evolution of the subhalo presented in \Figref{fig:innersubhaloexample} as a function of look back time. Top two sub-panels show radial distances of the object to the main branch of its $z=0$ host, in comoving units \& relative to host $R_{200\rho_c}$ respectively. Next two sub-panels show relative radial \& tangential velocities. Bottom two sub-panels show the object's $V_{\rm max}$ \& $M_{200\rho_c}$ evolution. Points are colour coded by radial distance from host. We also highlight points: squares indicate when the object is a subhalo of the host main branch, diamonds signify that the object is a subhalo of another halo, and stars indicate the object itself has $\geq20\%$ of its own mass in substructure. For all sub-panels we show the accretion time by a dashed vertical line. We also show several properties of host main branch by a dotted green line: $R_{200\rho_c}$ in the top sub-panel; scale radius in the $2^{\rm nd}$ sub-panel; $V_{\rm max}/10$ in the $5^{\rm th}$ sub-panel; and $M_{200\rho_c}/100$ in the $6^{\rm th}$ sub-panel. We also highlight when the host main branch is a subhalo or contains significant amounts of substructure by a diamond and star respectively.}
    \label{fig:subhaloorbit}
\end{figure}

The low density contrast might also suggests that this object is perhaps artificial, despite being identified by \textsc{VELOCIraptor}, \textsc{AHF}, and \textsc{rockstar}. To verify its physical origin, we must examine is history. We find that it is present in \textsc{hbt+} catalogue and thus must have originated from a 3DFOF halo. We show the mass accretion history as reconstructed by \textsc{TreeFrog} \cite[][]{treefrogpaper} from the \textsc{VELOCIraptor} catalogue along with its the radial motion, radial and tangential velocities and maximum circular velocity in \Figref{fig:subhaloorbit}, highlighting how well \textsc{VELOICraptor} works (see \Figref{fig:subhaloorbit2} in \Secref{sec:appendix:orbits} for more examples). 

\par
At $z=0$, this subhalo is found on a primarily radial orbit deep within the host. This object's first progenitor formed at a redshift of $z_{\rm form}=5.1$ with 32 particles and gradually moves closer to the main branch of the host halo. On its way, it experiences a significant merger event at high redshift, i.e., it contains a subhalo that has a mass ratio of $\geq1:10$ as indicated by the open diamond and open stars surrounding its track. This event also corresponds to when it experiences significant fluctuations in mass \& $V_{\rm max}$.  The fluctuations are quite severe, changing masses by factors of $\sim2$, as the object is not well resolved at this time, composed of $\sim200$ particles. The fluctuations in mass are also partially due to the fact that masses for subhalos are exclusive, whereas for field halos, the mass includes substructure. At these high redshifts, the main branch also experiences several major mergers, giving rise to mass fluctuations and changes in the relative motion of the subhalo to the host. 

\par 
Prior to its accretion, the object contains a single large substructure containing $\sim25\%$ of its total mass. The sudden drop in mass upon accretion is due to subhalo masses being exclusive in \textsc{VELOCIraptor}. Critically, the mass evolution after accretion is physically reasonable. Little mass is lost till pericentric passage, at which the system is shocked, increasing its $V_{\rm max}$ (and $R\left({V_{\rm max}}\right)$). After this impulsive heating, the halo begins to lose mass, the rate of mass loss decreasing as it reaches apocentre, which lies outside the halo at $2R_{200\rho_c}$. The object then plunges radially through the halo. The slight kinks in the radial and tangential velocities during this radial infall here are due to the host halo merging with a subhalo with a mass ratio of 1:6. The central regions of the main halo consist of two overlapping phase-space distributions with slightly different velocities that are rapidly phase-mixing. \textsc{VELOCIraptor} is no longer able to disentangle these objects, causing a small amount of jitter in the centre-of-mass velocity of the host.

\par
\begin{figure}
    \centering
    \includegraphics[width=0.45\textwidth,trim=1.cm 4.cm 1.25cm 28.5cm, clip=true]{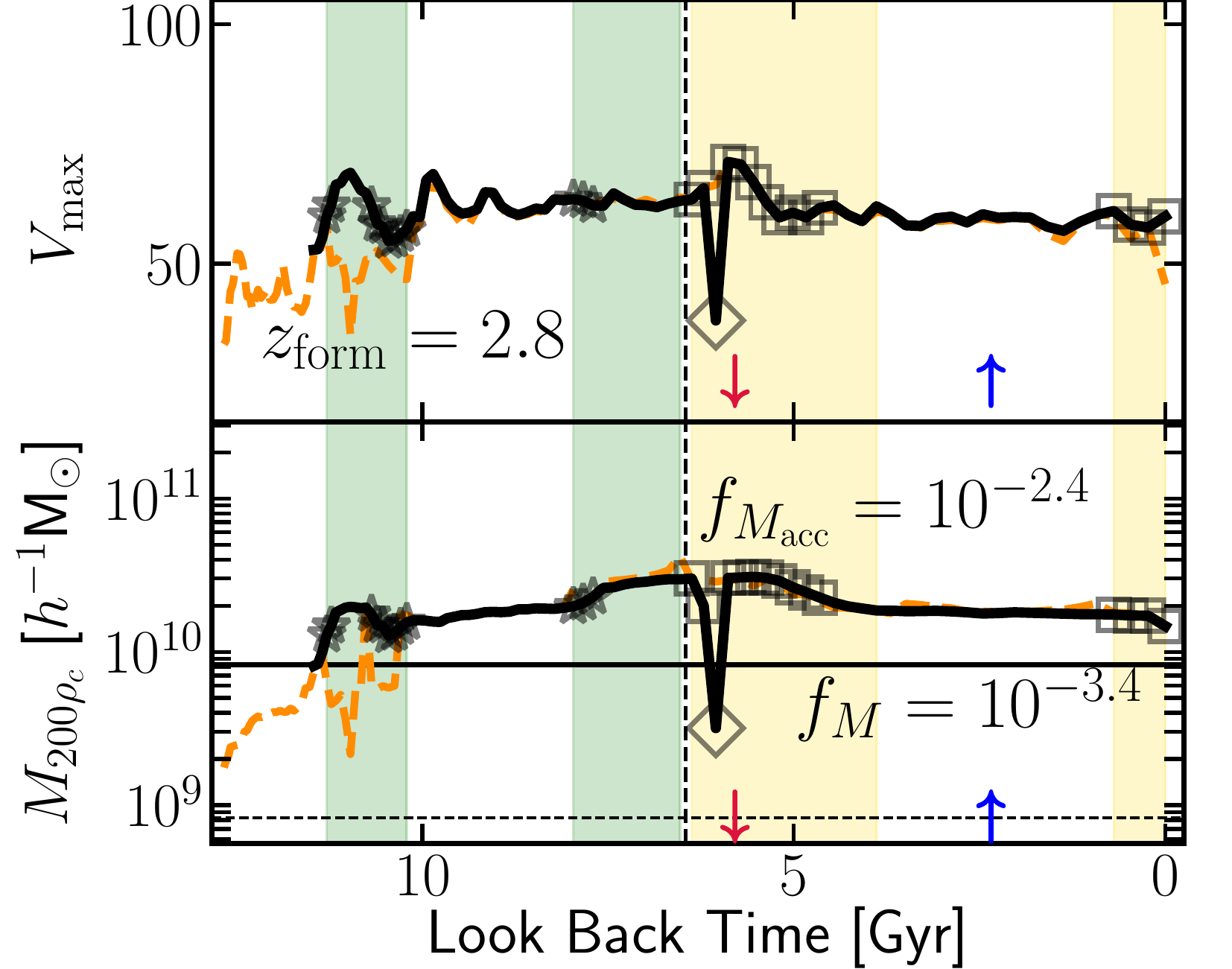}
    \caption{{\bf Reconstructed \textsc{AHF} Subhalo Evolution:} We plot the  $V_{\rm max}$ \& $M_{200\rho_c}$ evolution of the \textsc{ahf} counterpart to the subhalo presented in \Figref{fig:innersubhaloexample} as a function of look back time. We plot the \textsc{ahf} object with a solid black line, the \textsc{VELOCIraptor} object with a dashed orange line. Similar to \Figref{fig:subhaloorbit}, we highlight when the object is a subhalo of the host main branch, a subhalo of another halo, and when the object itself has $\geq20\%$ of its own mass in substructure. We also highlight periods when the \textsc{VELOCIraptor} object has significant substructure or is a subhalo by a shaded green and shaded yellow region respectively. We indicate when pericentric and apocentric passages occurs by $\downarrow$ \& $\uparrow$ respectively. For all sub-panels we show the accretion time by a dashed vertical line. .}
    \label{fig:subhaloorbitAHF}
\end{figure}

 
For comparison, we examine the counterpart identified by \textsc{ahf}, a configuration-based finder, which identifies a subhalo despite the low density contrast. The object is similar if lower mass at the last snapshot. As the orbital reconstructed orbital motion is similar, we focus on mass and maximum circular velocity evolution in \Figref{fig:subhaloorbitAHF}, highlighting where the object is a subhalo and has itself significant substructure. We also show the evolution of the \textsc{VELOCIraptor} object and highlight with shaded regions where the object contains significant substructure or is a subhalo. This figure shows that both codes recover similar mass evolution save for two key differences. The AHF subhalo experiences a rapid mass fluctuation in mass, dropping by an order of magnitude as the object approaches pericentre where density contrasts are low. The object also forms much later when composed of $\sim200$ particles, during a period where the object is undergoing a major merger. In the \textsc{ahf} catalogue, the object is lost for a few snapshots, truncating the halo merger tree. This figure indicates that in general both configuration-space and phase-space finders perform well and it is only during pericentric passages and major mergers that phase-space based finders outperform configuration-space based ones. 

\par
The other instance where a phase-space finder like \textsc{VELOCIraptor} outperforms configuration-space based ones is in recovering tidal debris. Tidal debris is not spatially overdense and requires measurement of the local velocity density. By using the local velocity density, \textsc{VELOCIraptor} naturally identifies a continuum of substructures from bound subhalos to unbound dynamically cold streams. This initial catalogue is cleaned by invoking an unbinding process. If we relax the unbinding criterion and also use the two stage iterative procedure described in \Secref{sec:velociraptor:substructures} to retain tidal features and debris, we have the structures shown in the bottom-right sub-panel of \Figref{fig:halodecomposition}, where we have limited the groups to those that have at most $50\%$ of their particle's bound. These objects consist of tidal shells originating from the larger merging subhalos and subhalos with large, extended tidal tails. For a thorough discussion of tidal debris, see \cite{elahi2013a}. Here we will focus on the recovered subhalo distribution. %

\subsection{Population}
\label{sec:results:velociraptor:pop}
\subsubsection{Halos}
\begin{figure}
    \centering
    \includegraphics[width=0.49\textwidth,trim=0.75cm 1.cm 0.75cm 0.75cm, clip=true]{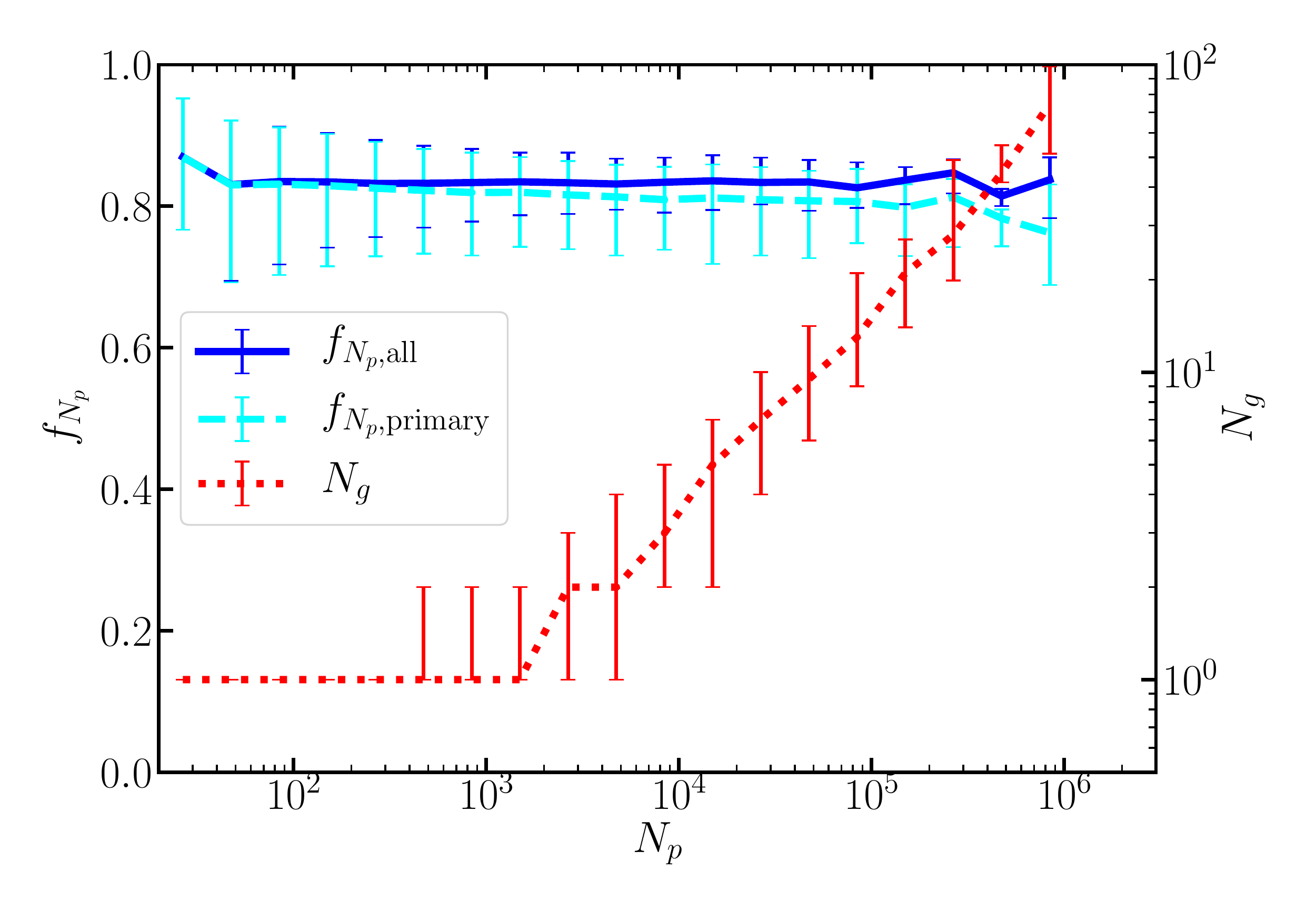}
    \caption{{\bf 6DFOF to 3DFOF stats:} we plot the fraction of particles in 6DFOF groups per 3DFOF group (blue solid), the fraction in the largest 6DFOF group (dashed cyan), and the number of 6DFOF groups per 3DFOF (right y-axis, red dotted line) as a function of the number of particles in the 3DFOF group. For each curve we plot the median, $16\%$ and $84\%$ quantiles.}
    \label{fig:3dto6dpop}
\end{figure}

We examine the impact of and the results from each stage of the algorithm using default parameters. We start by looking at halos identified with 3DFOF versus a 6DFOF. Using a 6DFOF does not significantly alter the input 3DFOF population as, on average, 3DFOF halos contain a 6DFOF object with $0.82^{+0.07}_{-0.10}$ of the mass of the original FOF object, independent of the number of particles in the 3DFOF as seen in \Figref{fig:3dto6dpop}. Consequently, the 6DFOF mass function should show a small suppression in mass relative to the 3DFOF mass function. The number of 6DFOF objects per 3DFOF group increases with increasing number of particles in the 3DFOF group as seen in \Figref{fig:3dto6dpop}. 

\par 
Due to resolution limits, not all 3DFOF objects contain a viable 6DFOF halo, particularly close to the imposed particle threshold of 20, where only $50\%$ of 3DFOF objects have a 6DFOF halo above this threshold. The absence of a 6DFOF halo in a 3DFOF object drops to $\lesssim1\%$ for 3DFOF objects composed of $\sim100$ particles. These objects are typically highly unrelaxed, unbound 3DFOF objects, i.e., spurious 3DFOF objects. 

\par
\begin{figure}
    \centering
    \includegraphics[width=0.49\textwidth,trim=0.75cm 1.cm 0.5cm 0.75cm, clip=true]{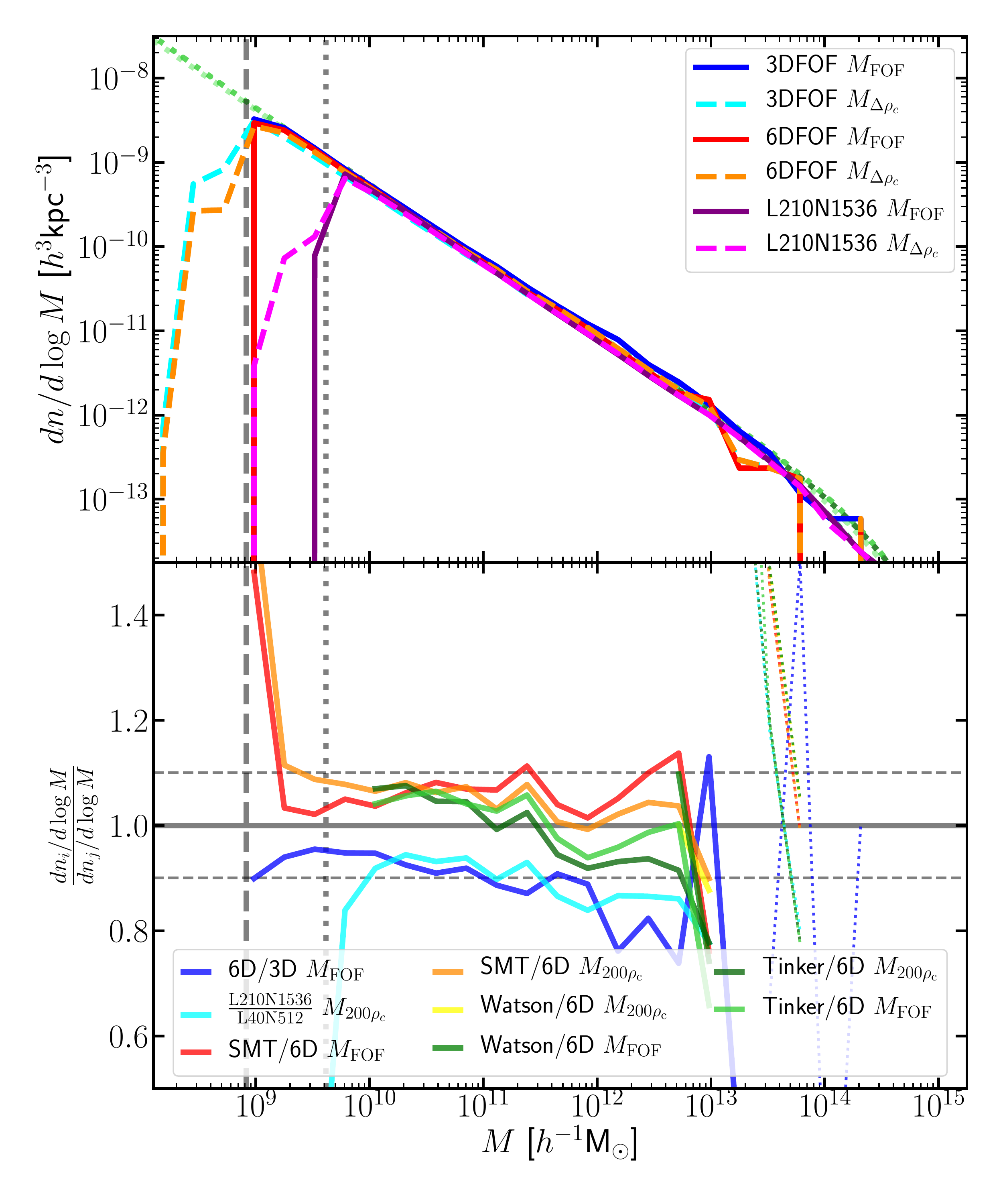}
    \caption{{\bf Halo Mass Functions:} We plot halo mass function measured using the 3DFOF and 6DFOF algorithm. The top panel shows the mass function along with several models, plotted as green coloured dashed lines. In the bottom panel we plot the radio of a interesting subset of results and models, with models calculated using {\sc HMFCalc} \cite[][]{hmfcalc}. Lines are thin at high masses when the number of halos in a given mass bin is below 10, i.e., the statistical variation exceeds $25\%$.}
    \label{fig:halomassfunc}
\end{figure}

 
The resulting halo mass from the different FOF algorithms and N-Body simulation are shown in \Figref{fig:halomassfunc}. For FOF masses, the 6DFOF mass function is effectively the 3DFOF mass function shifted to the left by $\approx0.1$~dex (as on average 6DFOF halos contain $80\%$ of the original 3DFOF halo's particles). The virial mass remains unchanged when comparing the 3DFOF halo to the largest 6DFOF object within the 3DFOF halo, with small mass differences due to small differences in the centre-of-mass. However, as there are on average 1.3 6DFOF groups per 3DFOF halo, the 6DFOF virial mass function has more halos per mass bin. The peak the virial mass distribution at low masses arises from loosely bound, poorly resolved halos with low overdensities. 

\par 
The residuals show that the 6DFOF mass function has fewer objects than the 3DFOF one at a given $M_{\rm FOF}$, as expected. We also compare the 6DFOF algorithm to three models, \cite{shethtormen2001}, \cite{tinker2010a} and \cite{watson2013a}. These models span a range of algorithms used to find halos and the type of mass recorded.  \cite{shethtormen2001} \& \cite{watson2013a} use 3DFOF algorithms, whereas \cite{tinker2010a} used to a spherical overdensity finder. \cite{watson2013a} uses $M_{\rm FOF}$, whereas the other two use $M_{\rm 200\rho_c}$. The 6DFOF relative to the models has fewer objects per mass bin. The systematic shift is of the same size as going from 3DFOF to 6DFOF. This is partially due to the 6DFOF naturally decomposing 3DFOF objects into dynamically distinct halos, although there are other systematic effects between the simulation and the theoretical curves arising from the finite volume of the box\footnote{The models are calibrated using larger volumes. The finite volume introduces systematic biases in mass functions, suppressing growth. Cosmic variance present in larger volumes is also absent.}. We also compare our reference simulation to our larger volume, lower mass resolution simulation, L210N1536. The simulations agree to within $\lesssim5\%$ for well resolved halo masses of $\gtrsim5\times10^{9}\Msunh$, though the larger volume simulation contains slightly fewer halos with $M_{200\rho_{c}}\lesssim10^{10.5}\Msunh$. 

\par 
The velocity function, not presented here for brevity, is effectively unchanged save for the fact that the 6DFOF is able to decompose 3DFOF objects into multiple halos, increasing the amplitude of the 6DFOF relative to the 3DFOF for well resolved halos with $V_{\rm max}\gtrsim30$~km/s.

\subsubsection{Subhalos}
\begin{figure}
    \centering
    \includegraphics[width=0.49\textwidth,trim=0.75cm 1.cm 0.5cm 0.75cm, clip=true]{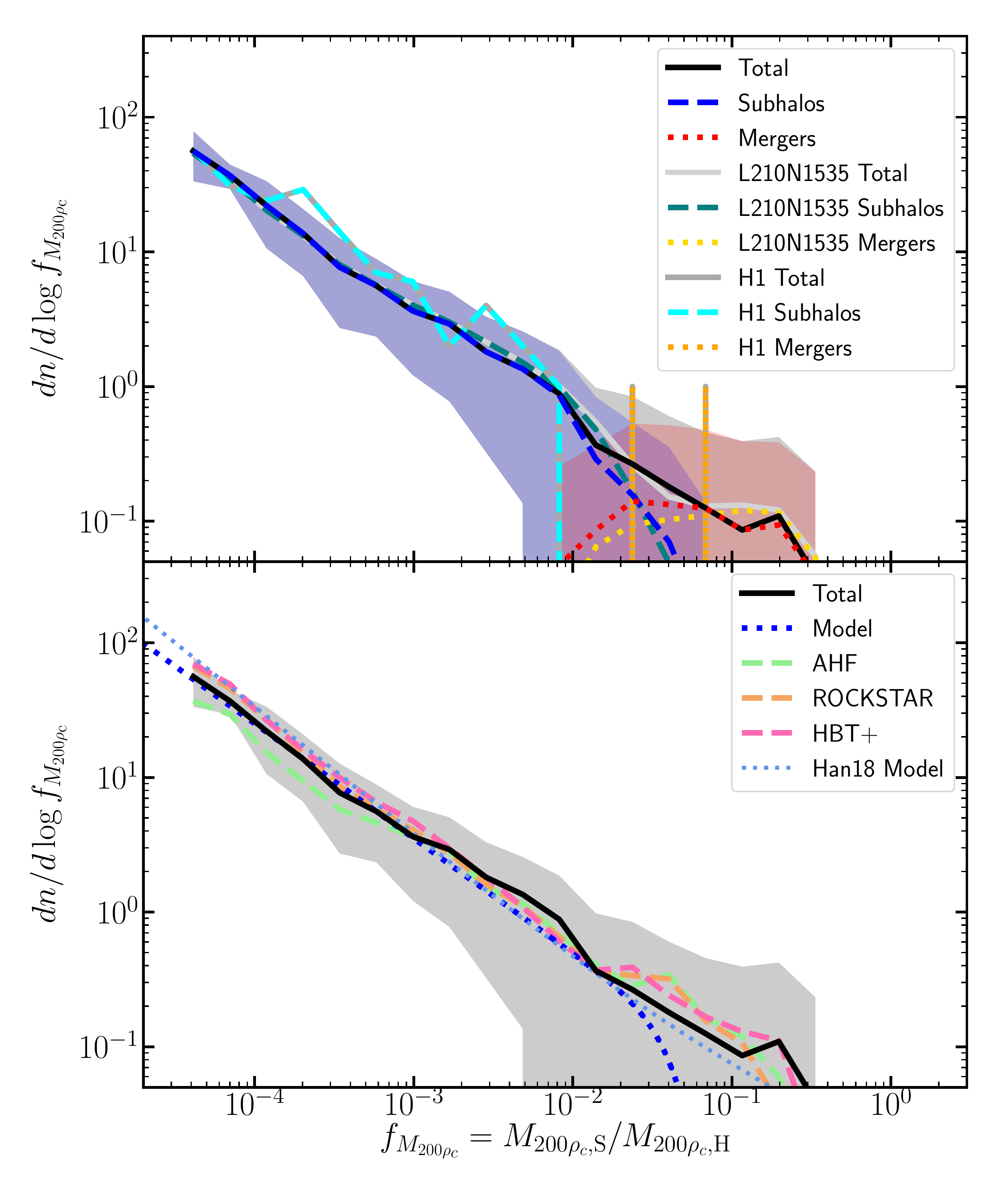}
    \caption{{\bf Subhalo Mass Function:} We plot the median subhalo mass function plus the $1\sigma$ scatter for all halos composed of $>=50000$ particles. We split the \textsc{VELOCIraptor} mass function into two categories, subhalos, and mergers. We also show the median distribution from a larger-volume, lower mass-resolution simulation L210N1536 and that from our fiducial example halo, H1. In the lower panel, for comparison, we show the power-law fit and the median distribution from \textsc{ahf}, \textsc{rockstar}, and \textsc{hbt+} using the L40N512 box, along with a best fit model and the model from \cite{han2018a}.}
    \label{fig:halosubhalomassfunc}
\end{figure}
\begin{figure}
    \centering
    \includegraphics[width=0.49\textwidth,trim=0.75cm 1.cm 0.5cm 0.75cm, clip=true]{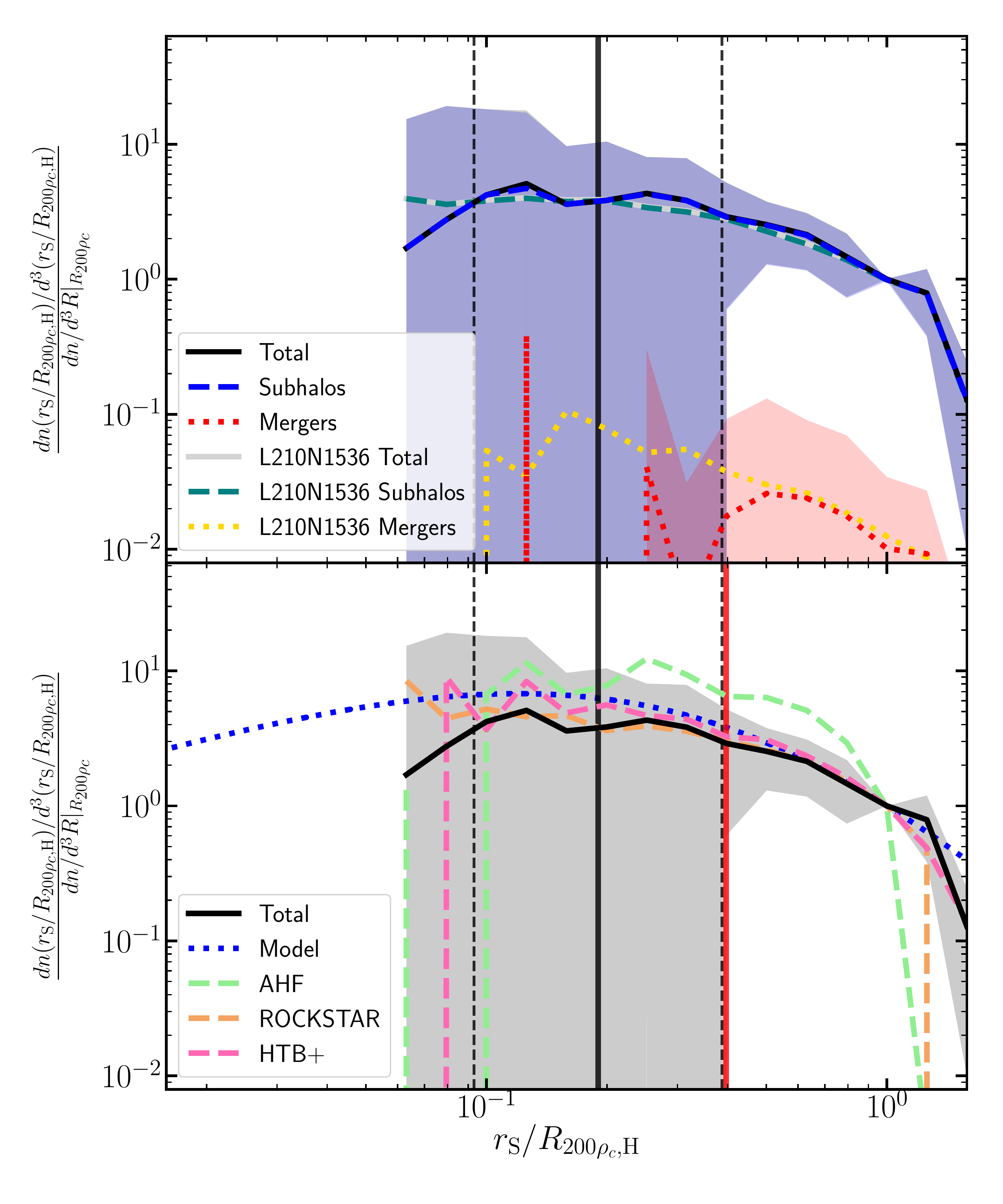}
    \caption{{\bf Subhalo Radial Distribution:} We plot the number density of subhalos. Similar to \Figref{fig:halosubhalomassfunc}, save we limit the analysis to halos composed of $>=10^{5}$ particles (so as to probe well inside the virial radius). The average scale radius and the $1\sigma$ scatter is shown by solid and dashed vertical lines respectively.}
    \label{fig:halosubhaloradialfunc}
\end{figure}
We next examine the results of subhalo/core search for our example halo and the population as a whole. To determine the average subhalo mass function we stack all halos composed of $\geq50000$ particles, i.e., all halos that at least probe the subhalo mass function down to masses fractions of $f_{M}\geq5\times10^{-4}$, with halo masses of $M_{200\rho_c}\gtrsim2\times10^{12}\Msunh$. There are $128$ such halos in our reference simulation. We focus on overdensity mass ratios $f_{M_{200\rho_c}}\equiv M_{200\rho_c{\rm S}}/M_{200\rho_c{\rm H}}$ presented in \Figref{fig:halosubhalomassfunc} (although using the total mass dynamically associated to a substructure relative to the dynamical mass exclusively associated with the parent subhalo, $f_{M_{\rm tot}}\equiv M_{\rm tot,S}/M_{\rm tot,H}$ does not significantly change the resulting mass function). For consistency across catalogs, we identify all objects with the virial radius of the host as subhalos. Here we classify substructures based on the specific method used to identify them to highlight any differences in the distribution arising from the methods: objects identified by the phase-space FOF algorithm on dynamically distinct particles (\Secref{sec:velociraptor:substructures}) are here referred to as subhalos; objects identified by searching for phase-space dense cores (\Secref{sec:velociraptor:mergers}) in the parent halo are classified as mergers (containing both major and minor merger events with mass ratios of $\gtrsim0.05$). This categorisation does not imply a hard physical difference between objects, it is simply to highlight any algorithmic differences. We also classify objects identified within substructures, so-called subsubstructures as subhalos. 

\par
On average, halos contain a total of $208_{-130}^{+22}$ subhalos with overdensity masses of $f_{M_{200\rho_c}}\gtrsim7\times10^{-5}$ (with the numbers increasing if looking at total dynamical masses with the same limit of $f_{M_{\rm tot}}\gtrsim7\times10^{-5}$ to $272^{+61}_{-177}$). Halos contain at least 1 subhalo with a mass of $f_{M_{200\rho_c}}\sim10^{-2}$. Significant merger events are not uncommon, with an average number per halo of $1.7\pm1.6$. The example halo contains subhalos with $10^{-5}\lesssim f_{M_{200\rho_c}}\lesssim10^{-2}$ and contains three large merger remnants with mass fractions of $f_{M_{200\rho_c}}\gtrsim4\times10^{-2}$. The fiducial halo has more substructure than the average (it lies close to the $+1\sigma$ envelop), which is to not unexpected given the number of merger remnants it contains and the fact that this 6DFOF halo lies at the nexus of three large merging halos (see \Figref{fig:halodecomposition}).

\par 
The median and halo-to-halo scatter seen in our small volume simulation is in agreement with that seen in our large volume, lower-mass resolution simulation, L210N1536, when applying the same particle number threshold (for clarity we only show the median for this simulation). The median distribution from L210N1536 is based on $3000$ halos with a higher median masses of $M_{200\rho_c}\approx3\times10^{13}\Msunh$. The agreement between different host halo masses indicates a mostly scale-free subhalo mass function. 

\par 
For comparison, we also show the results from \textsc{ahf}, a configuration-space based halo finder, \textsc{rockstar}, a phase-space halo finder, and \textsc{hbt+}, a 3DFOF tracker. These mass functions agree within the scatter modulo differences in the definition of subhalo masses, which vary from halo finder to halo finder \cite[see][for a discussion of mass definitions]{knebe2013a}. \textsc{VELOCIraptor} can report a variety of different masses: bound mass, total dynamical mass, overdensity masses, etc. The first two masses are calculated using a exclusive particle list. For halos, it calculates inclusive spherical overdensity masses. For subhalos, all these masses are calculated based on a list of particles belonging exclusively to the object, neglecting the background host and internal substructures. \textsc{rockstar} also calculates masses for subhalos in a similar fashion using exclusive particle lists. \textsc{ahf} calculates inclusive spherical overdensity masses defined by a saddle point and processed through an unbinding procedure, so most resembles the spherical overdensity masses of \textsc{VELOCIraptor} and \textsc{rockstar}. \textsc{hbt+} returns bound masses based on the initial FOF envelop and does not allow subhalos to accrete mass from their surrounding host halo, though they can accrete material from subsubhalos, those objects that were subhalos when the object itself was a FOF halo. This mass best corresponds to the total bound mass calculated by \textsc{VELOCIraptor}. 

\par 
Although the mass functions agree, there are systematic differences in the number of subhalos per halo found by each finder. Given the high cadence of the input 3DFOF catalogue\footnote{With high cadence, a 3DFOF tracker is unlikely to miss the formation of a halo.}, \textsc{hbt+} is a useful reference catalogue. \textsc{VELOCIraptor} finds similar numbers of objects composed of $\geq20$ particles within $R_{200\rho_c}$ of large halos as \textsc{hbt+}, identifying $98\pm7\%$ of all 3DFOF halos tracked, some of the variation due to differences in the centre-of-mass. \textsc{ahf} finds a slightly smaller percentage of $84\pm10\%$, the lower number arising from small, low-density subhalos. The outlier is \textsc{rockstar}, which identifies a factor of $1.85^{+0.15}_{-0.2}$ more objects, though a significant fraction appear to be diffuse, possibly spurious, phase-space structures with low $M_{200\rho_c}$, with some never reaching overdensities of $200\rho_c$. Removing these low density objects from the halo catalogue places it more in line with the other codes, though it still identifies a factor of  $1.05^{+0.1}_{-0.05}$ more objects than \textsc{hbt+}. 

\par
The average subhalo distribution is well-characterised by a power-law with an exponential dampening at the high mass. We fit the average mass function using \textsc{emcee}\cite[][]{emcee} with
\begin{align}
    dn/df_M=Af_M^{-\alpha}\exp\left[-\left(f_{M}/f_o\right)^\beta\right], 
\end{align}
focusing on subhalos explicitly (that is those objects identified by the method outlined in \Secref{sec:velociraptor:substructures} with typical mass ratios of $f_M\lesssim10^{-2}$), and ignore minor/major merger remnants (objects identified by the method outlined in \Secref{sec:velociraptor:mergers} with typical mass ratios of $f_M\gtrsim10^{-2}$). We find $\log A=-1.7_{-1.0}^{+0.7}$, $\alpha=1.85_{-0.18}^{+0.16}$, $\log f_o=-1.33\pm0.9$, $\beta=3.2\pm1.9$ for $M_{200\rho_c}$, though the fit does not vary drastically if we use total masses. The amplitude and power-law are consistent with previous studies  \cite[e.g.][]{madau2008,springel2008,stadel2008,gao2012,onions2012,rodriguezpuebla2016a,jiang2016b,han2018a}. The scale of the exponential dampening occurs at $f_M\approx0.05$, in agreement with recent studies \cite[e.g.,]{jiang2016b,han2018a}. 

\par 
Large mass ratio objects, that is minor and major mergers, appear to be characterised by a skewed-Gaussian-like distribution. The fact that mergers follow a different distribution than subhalos is not surprising as once objects are large enough, they become less prone to tidal stripping and more affected by dynamical friction. Given number of merger remnants in this data set, we refrain from fitting the distribution, though the average of $\log f_{M_{200\rho_c},{\rm mergers}}=-1.2\pm0.8$ is in agreement with \cite{elahi2018a}, who used a larger data set to fit results. We find that the total subhalo mass function also agrees with the double power-law fit in \cite{han2018a}, although the the second power-law describing the high mass end is poorly constrained with values of $1.1-1.5$ (for completeness we show the double power-law from \cite{han2018a} in the figure). 

\par 
The fact that the total subhalo mass function (subhalos+mergers) is not characterised by a single power-law is also seen by \cite{han2018a}, (see also \textsc{hbt+} in \Figref{fig:halosubhalomassfunc}). They argued for characterising the subhalo mass function by a double Schetcher function with a steep power-law for low mass fractions and a flatter that dominates at high mass fractions. Given the small number of large subhalos, which also span a very small range in $f_M$, it is difficult to differentiate between either model with the number of host halos in this sample and the halo-to-halo scatter. 

\par
The radial distribution of subhalos in the form of the differential number density $dn/dV$ normalised by the number of objects at the virial radius is shown in \Figref{fig:halosubhaloradialfunc}. We limit our sample to halos composed of $\geq10^{5}$, as these halos contain significant amounts of substructure and have density profile converged to radii of $\approx10^{-2}R_{200\rho_c}$ \cite[][]{power2003}. 

\par 
We fit a generalized NFW-like profile to the distribution:
\begin{align}
    dn/dV\propto\left(r/r_s\right)^{-\alpha}\left(1+r/r_s\right)^{-\beta},
\end{align}
where $r_s$ is the scale radius, and $\alpha$ and $\beta$ represent the inner and outer slopes. This fit is motivated by the fact that halo dark matter density profiles follow this profile with $\alpha=1$ and $\beta=2$. Subhalos should be radially distributed in a similar to the smoothly accreted dark matter. We find an optimal fit of $r_s=0.4\pm0.1R_{200\rho_c}$,  $\alpha=0.10\pm0.23$ and $\beta=3.85^{+0.11}_{-0.23}$. This profile that is flatter than a halo density profile in the inner regions, in agreement with previous studies \cite[see for instance][where they find an inner slope of $\sim0.3$ for a very well resolved $10^{12}\Msunh$ halo, a fit that well describes halos over a wide range in masses.]{han2016a}, although our halos are not well resolved enough precisely constrain the exact slope of the inner profile. Only very high resolution zoom simulations, such as Aquarius \cite{springel2008}, contain enough subhalos to properly constrain the inner slope and even then, since subhalos spend most of their time at apocentre and not pericentre, few subhalos are present in the very central regions for long. 

\par 
The second power-law index implies that the logarithmic slope $d\ln n/d\ln r=-\alpha-\beta\left(r/r_s\right)/\left(1+r/r_s\right)$ is steeper than a NFW profile and even our fit does not capture the steep slope of the subhalo distribution. However, we stress that at the virial radius, both the subhalo radial distribution and the halo density profiles have similar logarithmic slopes of $\sim-2.8$. Only at even larger radii do subalos drop off faster than an NFW profile\footnote{It should be noted that average density profiles of cluster mass halos also fall off faster than an NFW profile for $1\lesssim R_{200\rho_c}\lesssim1.6$ before becoming flatter than an NFW profile at larger radii \cite[e.g.][]{diemer2014a}.}. 


\par 
Comparing results, we find that the median distribution from the larger volume, lower-mass resolution simulation agrees with our L40N512. The fact that host halos in the L210N1536 sample are $\sim10$ times more massive argues in favour of a scale-free radial distribution. The \textsc{ahf} radial distribution is biased to larger radii and contains fewer subhalos deep within the host. The lack of subhalos within $0.1R_{200\rho_c}$ has to do with the configuration-based nature of \textsc{ahf}. Density contrasts between subhalo and host are small, making it more difficult to separate subhalos from the background. Both \textsc{hbt+} and \textsc{rockstar} agree within the halo-to-halo scatter.

\par
\begin{figure}
    \centering
    \includegraphics[width=0.49\textwidth,trim=0.75cm 1.cm 0.5cm 0.75cm, clip=true]{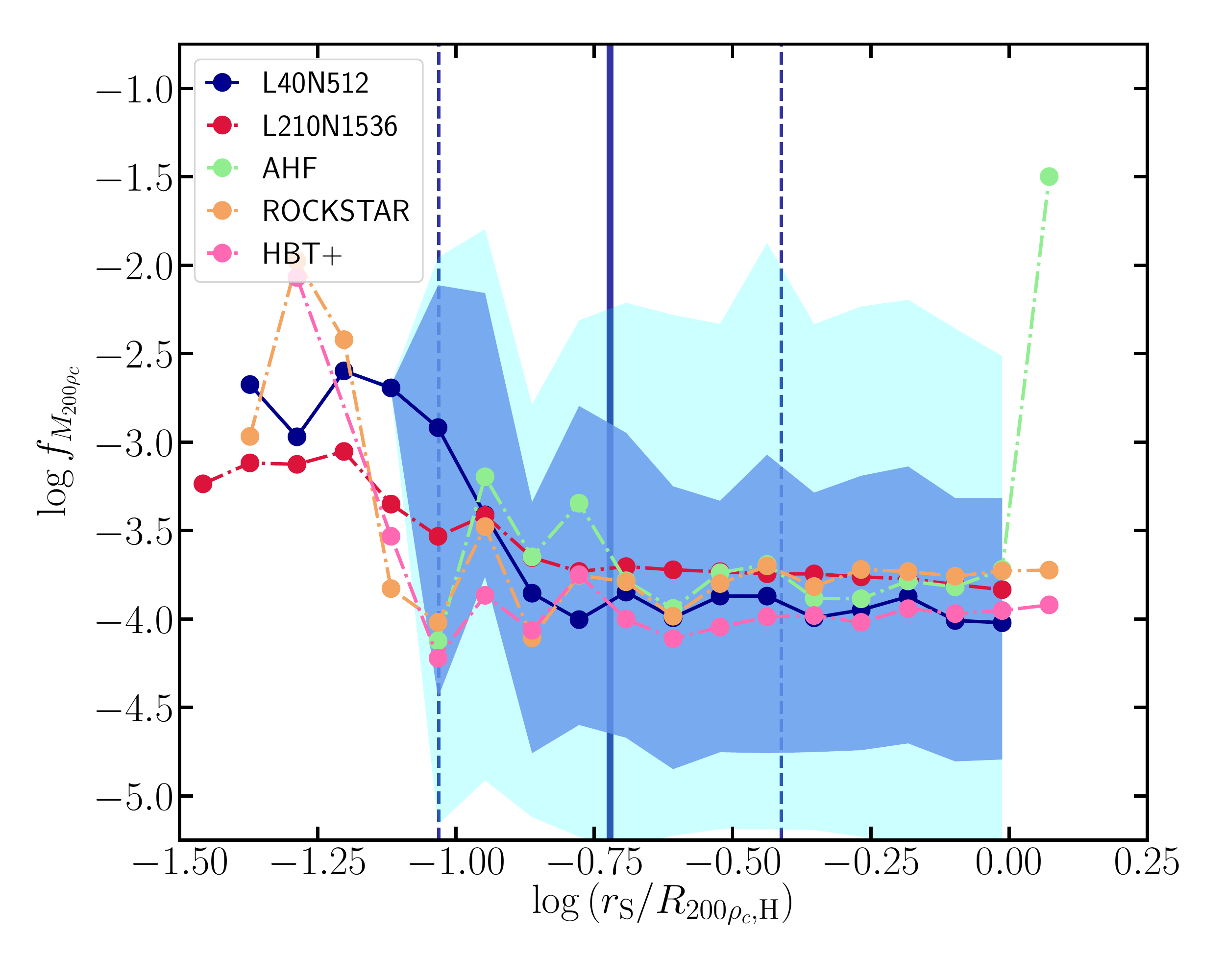}
    \caption{{\bf Subhalo Radial-Mass Distribution:} We plot median subhalo mass at a given radius as a solid blue line, along with the $16,84$ and $2.5,97.5$ quantiles as filled blue and cyan regions. The average scale radius and the $1\sigma$ scatter of host halos is shown by solid and dashed vertical lines respectively. We also show the median distribution for our L210N1536 run, \textsc{ahf}, \textsc{rockstar}, and \textsc{hbt+}.}
    \label{fig:halosubhaloradialmassdep}
\end{figure}

 
We now focus on mass of subhalos as a function of radius, where here we identify all objects within the virial radius of the host. The average substructure radial-mass dependence is shown in \Figref{fig:halosubhaloradialmassdep}, where we again stack all well-resolved halos, scaling subhalo masses and radial distances by virial masses and radius of the parent halo. The mass distribution for most radial bins, both in the median and the scatter, shows little radial dependence. The total population shows a very weak correlation with the Pearson covariance coefficient of $0.1\pm0.3$, which is consistent with no dependence. 

\par
Only the inner radii, typically within the scale radius of the host parent, do subhalo masses strongly depend on radii. The median subhalo mass markedly increases in the central regions. There are even subhalos with mass ratios as large as $f_{M_{200\rho_c}}\sim0.2$ found within $0.27R_{200\rho_c}$. This radial-mass dependence is also present in our larger volume, lower mass resolution run. The reason for this trend is two-fold: 1) large subhalos are strongly affected by  dynamical friction, pulling both their pericentres and apocenters inward; 2) large subhalos are also less prone to tidal disruption. Thus we should expect the inner regions to be dominated by large subhalos. 

\par 
This trend is also seen in \textsc{hbt+}. By tracking 3DFOF halos, \cite{han2018a} found the inner regions of halos contain large subhalos that remain trapped due to dynamical friction. \textsc{rockstar}, another phase-space finder also reproduces the general trend \footnote{As mentioned previously, the  \textsc{rockstar} catalogue contains low physical density objects with $M_200\rho_c$ masses below the particle number threshold used, with some having objects having densities below $200\rho_c$, i.e., $M_{200\rho_c}=0$. Here we limit the catalogue to objects with $M_{200\rho_c}\geq20m_p$, where $m_p$ is the particle mass.}. In contrast, configuration-space based finders like \textsc{ahf} shows a bias in the opposite direction in the very inner regions, and has no subhalos with $f_{M_{200\rho_c}}\sim0.2$ within $\gtrsim0.4R_{200\rho_c}$. 

\par
\begin{figure}
    \centering
    \includegraphics[width=0.49\textwidth,trim=0.75cm 1.cm 0.5cm 0.75cm, clip=true]{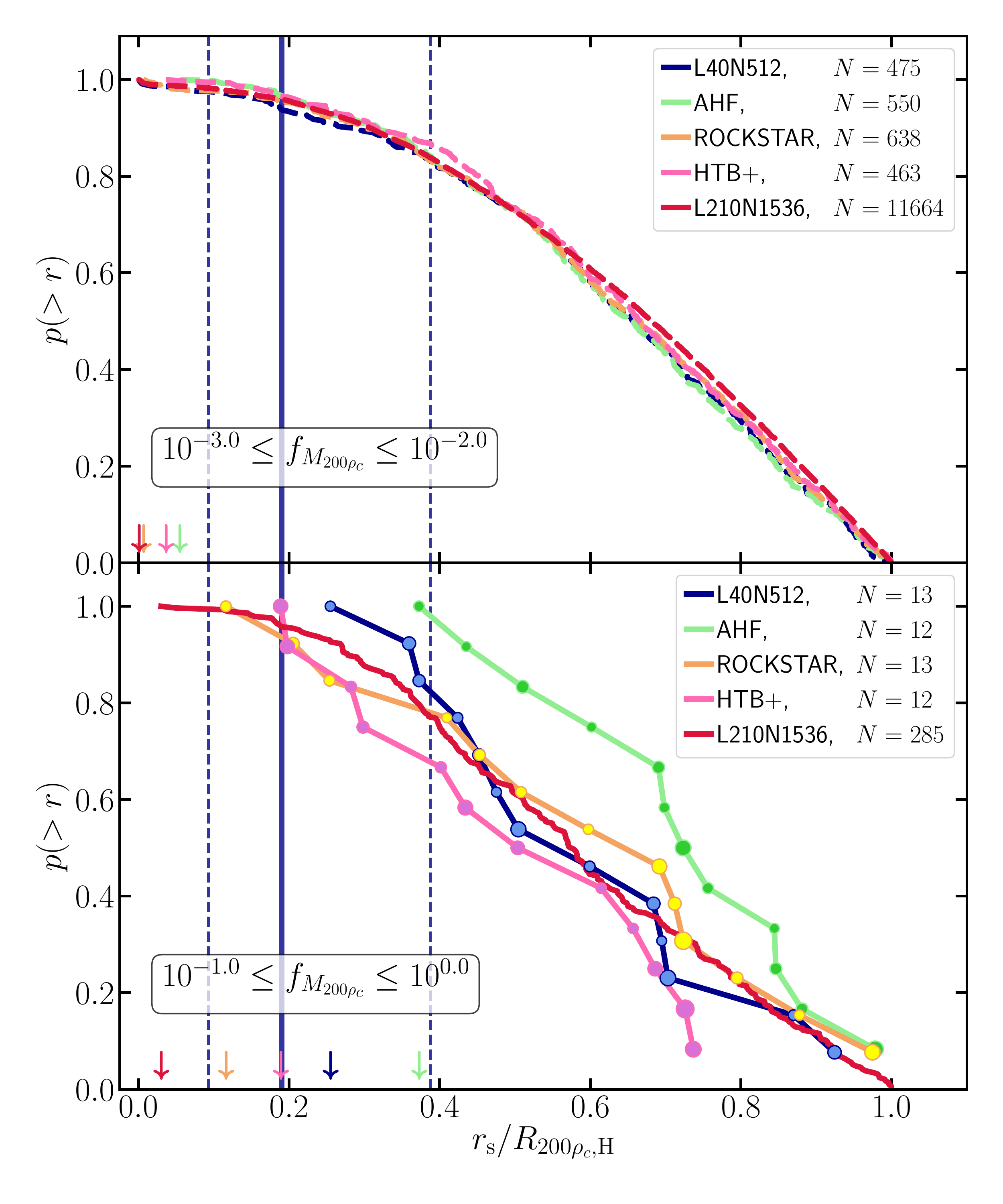}
    \caption{{\bf Subhalo Radial with Mass Cuts:} We plot the normalised cumulative number distribution of subhalos in two mass fraction bins containing low and high mass subhalos (in top and bottom panels). We emphasise the inner most subhalo with an arrow and also show the number of subhalos in each bin. For the lower panel, we also plot a circle scaled by the mass of the subhalo for each save the \textsc{VELOCIraptor} results for our larger L210N1536 simulation. We also show the average scale radius and the $1\sigma$ scatter of host halos by solid and dashed vertical lines respectively.}
    \label{fig:halosubhaloradialmassdepmasscuts}
\end{figure}

 
To further investigate differences between codes, we compare the normalised cumulative radial distribution of subhalos in \Figref{fig:halosubhaloradialmassdepmasscuts} to further examine this apparent radial-mass dependence, focusing on low and high mass subhalos. Our lower mass bin samples halos composed of $\sim100-1000$ particles for the smallest halos in this sample, well above the particle number threshold used to identify structures. Our upper mass bin effectively chooses major mergers. We find little difference between codes, with the inner most objects found well within the scale radius of the host halo. There is greater disagreement for large subhalos, in part owing to how the centre of a halo is defined (most bound particle, shrinking spheres estimate of mass, total bulk centre of mass). Nevertheless, we see that \textsc{ahf} is noticeably more biased to identifying large subhalos at larger radii that the other codes, a consequence of its configuration-space based approach. 

\section{Discussion and Conclusion}
\label{sec:discussion}
We have presented \textsc{VELOCIraptor}, a novel code designed to identify halos, subhalos, tidal debris and galaxies in both N-body and full hydrodynamical simulations using phase-space information. We have demonstrated that the code robustly identify (sub)halos, particularly cases that are typically notoriously difficult for such codes, namely the mass reconstruction of subhalos deep inside their host halo and major mergers. We summarise key features/results below.

\par
\textsc{VELOCIraptor} identifies structures in a multi-step process. For N-body simulations, it first identifies field halos using a 3DFOF followed by a 6DFOF algorithm. The next step identifies substructure in each halo in two stages. The first stage uses the previously developed algorithm described in \cite{elahi2011}, finding velocity outliers (so-called peaks above the Maxwellian sea) and linking particles using a phase-space FOF. The next stage is to find any remaining large minor/major mergers using an iterative search for dense phase-space cores that are then grown in an iterative fashion using phase-space tensors. 

\par 
We find that 6DFOF objects are more representative of dark matter halos than 3DFOF objects as 3DFOF objects can link separate virialised overdensities together via particle bridges. The 6DFOF step separates early stage accretion/merger events, with the average number of 1.3 6DFOF objects per 3DFOF objects. The 6DFOF also removes outer unbound particles from the 3DFOF candidate, with FOF masses changing by $M_{\rm 6DFOF}=0.82M_{\rm 3DFOF}$ while leaving spherical overdensity masses, particularly $200\rho_c$, unchanged. 

\par 
The substructure algorithm \cite[tested in][and shown to identify both subhalos and tidal debris]{elahi2011, elahi2013a} has the advantage over other algorithms of being able to identify subhalos deep within a host halo, where density contrasts relative to background are negligible. We highlighted a particular example where the average logarithmic density contrast between the subhalo and the host halo are $\sim1$, yet its particles are very distinct in velocity space. This subhalo does not undergo rapid artificial decrease in mass that affects most subhalo configuration-space based finders. 

\par
The merger algorithm, a new addition to the code, is fully described here. This algorithm uses full phase-space tensors to assign particles to any phase-space dense cores that are not already tagged as substructure. This technique, inspired by \textsc{rockstar} \cite[][]{rockstar} and Gaussian mixture models, can separate substructures from the main halo deep within the host (at least up to the scale radius of a host halo). The use of phase-space tensors allows for the mass assignment scheme to asymmetric tidal features associated with an object, unlike \textsc{rockstar}, which uses a scalar dispersion to assign particles. The iterative growth is also more physical than assigning particles using Gaussian mixture models, which assume a global dispersion tensor. This method does not necessarily artificially shrink halos as they move towards pericentre, as seen in the example figures in the appendix, though the scheme can occasionally lose halos or result in mass fluctuations of a few when objects overlap significantly. This can be alleviated somewhat by using a finer steps when searching for cores and assigning mass. 

\par
The resulting subhalo mass function reproduces the mass and radial distribution seen in codes that track particles, such as \textsc{hbt+}. Like this FOF tracker, the subhalo mass function can be decomposed into a distribution for low and high mass ratios. The low mass ratio end is described by a power-law with an exponential cut-off, with an index of $\alpha=1.85_{-0.18}^{+0.16}$ and a cut-off mass ratio scale of $f_o\sim0.05$. Our simulation does not have enough halos to well constrain the high mass end it can either be characterised by a power-law with a much flatter slope or possibly a lognormal distribution in mass ratio. 

\par
Critically, \textsc{VELOCIraptor} can recover the radial-mass distribution seen in tracking codes like \textsc{hbt+}, with larger subhalos found at smaller radii, {\em{without the need of tracking}}. The central regions within the scale radius of a halo are dominated by large subhalos and merger remnants. Although our fiducial simulation only contains a small sample of $\sim50$ well resolved halos composed of $\gtrsim10^5$ particles, which is not enough to rigorously constrain the inner radial distribution, these halos are resolved enough for this trend to be observed by \textsc{hbt+} and recovered by \textsc{VELOCIraptor}. This is in contrast to the distribution recovered by configuration-space based finders. The code also does not introduce possibly spurious phase-space structures like \textsc{rockstar}, which also recovers the radial-mass dependence. 

\par 
This radial-mass dependence is seen in our larger volume simulation, which contains $\sim1500$ well-resolved halos, including $\sim 50$ halos composed of $10^{6}$ particles. As we do not analyse this simulation with \textsc{hbt+}, we cannot definitively say that the observed trend is that recovered by tracking, though given the results from our fiducial simulation, it is likely in agreement. 

\par 
The code is in active development. New input interfaces for hydrodynamical simulations are being developed \cite[e.g.][]{canas2018a} and it is being incorporated into the \textsc{swift} Hydrodynamical N-Body code \cite[\href{http://www.swiftsim.com}{\url{www.swiftsim.com}}][]{swiftsimcode}. Additional libraries are being integrated to improve the parallel efficiency, such as the ADIOS library, designed for parallel IO at the $\sim10^{4}$ node scale, and METIS for efficient MPI decomposition. The output produced also lends itself to large-scale processing as it produces compressed, self-describing binary HDF5 data.

\par 
Finally \textsc{VELOCIraptor} is not limited to analysing cosmological simulations. The primary substructure algorithm is suited to finding clustering in a variety of data. One novel application could be to decompose data from GAIA \cite[][]{gaia2}, which contains 5-dimensional phase-space information for 1.3 billions stars, and full 6D phase-space information for 7 million in the Milky Way. Early analysis shows the mean velocity structure of the Milky Way disk is complex, with features indicative of substructure in the solar neighbourhood \cite{gaia2018a}. This data set is only just beginning to be mined for kinematic structures \cite[e.g.][]{hawkins2018a,price-whelan2018a,marchetti2018a}. For instance, \cite{castro-ginard2018a} used clustering algorithms and artificial neural networks to identify open clusters in the GAIA data set. This method essentially looks for full phase-space (configuration and velocity) clustering akin to a 6DFOF algorithm, as such is tailored to identifying open clusters. The nature of the substructure algorithm in \textsc{VELOCIraptor} makes it well suited for identifying open clusters and other substructures and even be extended to use other information, such as metallicity, making analysing this data with the code an interesting exercise. 

\begin{acknowledgements}
The authors would like to thank the anonymous referee for insightful comments that helped improve the clarity of the text. 

\par 
RC is supported by the SIRF awarded by the University of Western Australia Scholarships Committee, and the Consejo Nacional de Ciencia y Tecnolog\'ia (CONACyT) scholarship No. 438594 and the MERAC Foundation. RP is supported by a University of Western Australia Scholarship. Parts of this research were conducted by the Australian Research Council Centre of Excellence for All Sky Astrophysics in 3 Dimensions (ASTRO 3D), through project number CE170100013. CL is funded by a Discovery Early Career Researcher Award DE150100618. CL also thanks the MERAC Foundation for a Postdoctoral Research Award. 

\par
The authors contributed to this paper in the following ways: PJE ran simulations and analysed the data, made the plots and wrote the bulk of the paper. PJE is the primary developer of both {\sc VELOCIraptor}. RC, RT \& JW designed and developed of various aspects of the code: RC developed the core search; RT developed the compilation infrastructure; and JW bug tested and developed the interface with \textsc{swiftsim}. RP, CL, CP, \& AR assisted in the design of various aspects of the code. All authors have read and commented on the paper.

\paragraph*{Facilities} Magnus (Pawsey Supercomputing Centre)
\paragraph*{Software} 
\begin{itemize}
    \item \textsc{VELOCIraptor} \href{https://github.com/pelahi/VELOCIraptor-STF}{\url{https://github.com/pelahi/VELOCIraptor-STF}}
    \item \textsc{TreeFrog} \href{https://github.com/pelahi/TreeFrog}{\url{https://github.com/pelahi/TreeFrog}}
    \item \textsc{NBodylib} \href{https://github.com/pelahi/NBodylib}{\url{https://github.com/pelahi/NBodylib}}
    \item \textsc{VELOCIraptor\_Python\_Tools} \href{https://github.com/pelahi/VELOCIraptor_Python_Tools}{\url{https://github.com/pelahi/VELOCIraptor_Python_Tools}}
    \item \textsc{MergerTreeDendograms} \href{https://github.com/rhyspoulton/MergerTree-Dendograms/}{\url{https://github.com/rhyspoulton/MergerTree-Dendograms}}
    \item \textsc{ahf} \href{http://popia.ft.uam.es/AHF/Download.html}{\url{http://popia.ft.uam.es/AHF/Download.html}}
    \item \textsc{rockstar} \href{https://bitbucket.org/gfcstanford/rockstar}{\url{https://bitbucket.org/gfcstanford/rockstar}}
    \item \textsc{hbt+} \href{https://github.com/Kambrian/HBTplus}{\url{https://github.com/Kambrian/HBTplus}}
\end{itemize}
\paragraph*{Additional Software} Python, Matplotlib \cite[][]{matplotlib}, Scipy \cite[][]{scipy}, emcee \cite[][]{emcee}, SciKit \cite[][]{scikit}, Gadget \cite[][]{gadget2}
\end{acknowledgements}

\bibliographystyle{pasa-mnras}
\bibliography{velociraptor.bbl}

\begin{appendix}

\section{Orbits}
\begin{figure*}
    \centering
    \includegraphics[width=0.45\textwidth,trim=1.cm 4.cm 1.25cm 5.5cm, clip=true]{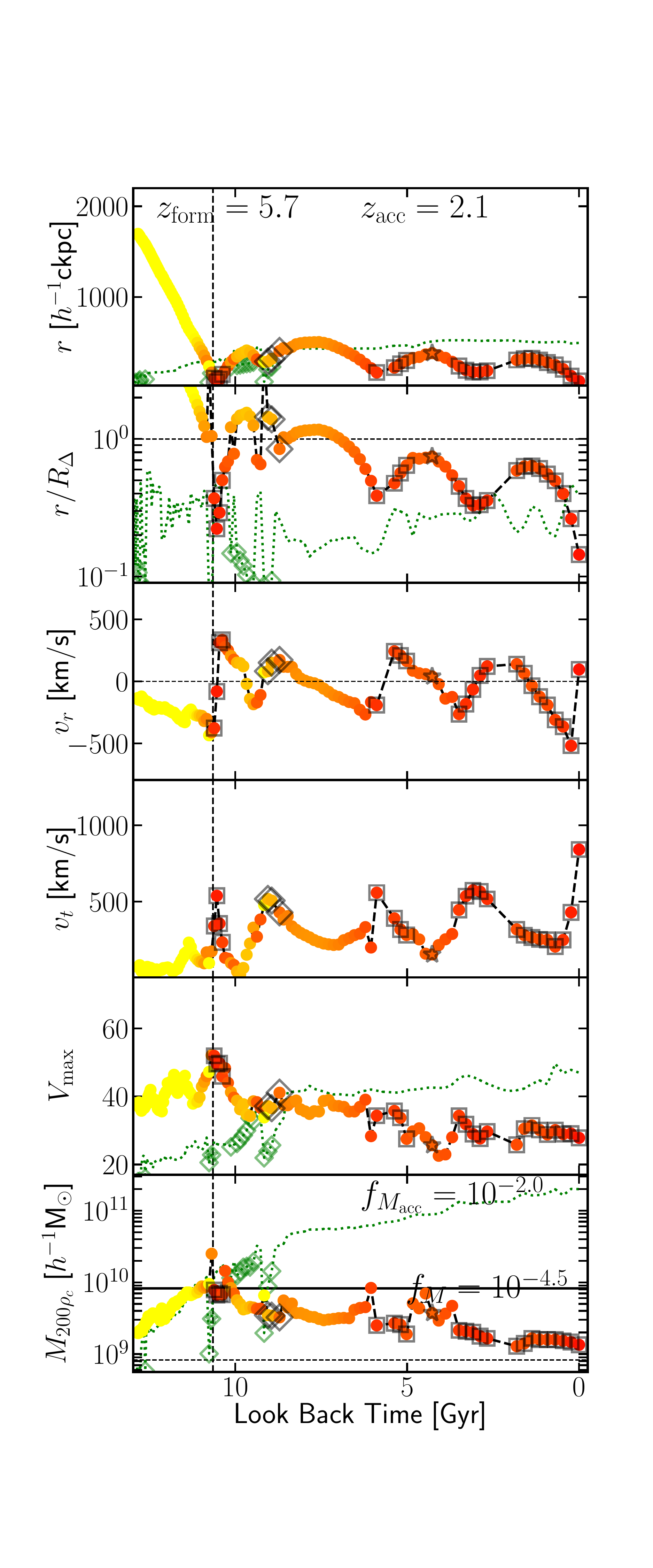}
    \includegraphics[width=0.45\textwidth,trim=1.cm 4.cm 1.25cm 5.5cm, clip=true]{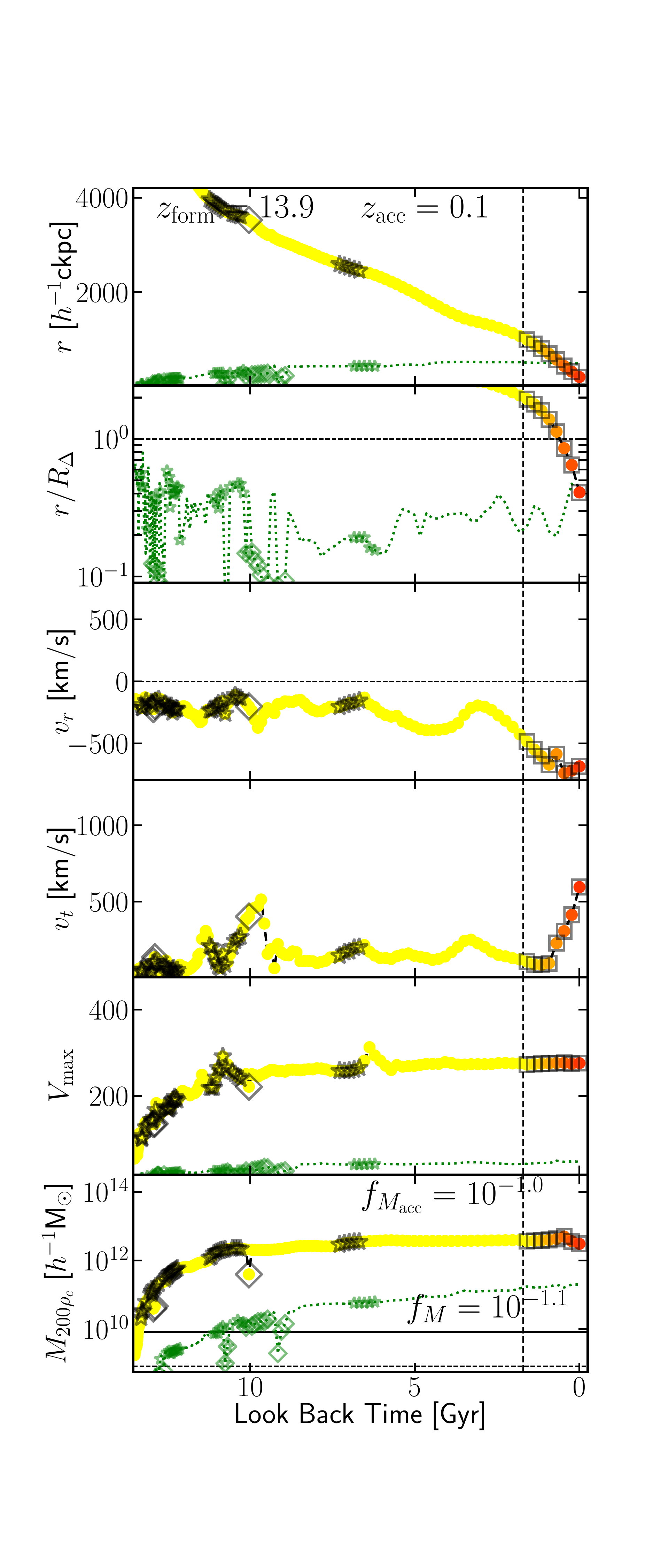}
    \caption{{\bf Reconstructed Subhalo Orbital \& Evolution:} We plot the orbital life of a poorly-resolved subhalo found at $r/R_{200\rho_c}=0.15$ (left) and a large-subhalo found at $r/R_{200\rho_c}=0.41$. Similar to \Figref{fig:subhaloorbit}.}
    \label{fig:subhaloorbit2}
\end{figure*}

We show the orbits of a low mass subhalo accreted at high redshift and large subhalo accreted at late times in \Figref{fig:subhaloorbit2}. The poorly resolved subhalo is still recovered when deep inside the host halo even when composed of $\sim30$ particles. There are gaps in the subhalo's orbit where it is momentarily lost. The large subhalo is accreted at late times and is still approaching pericentre. It does lose an appreciable amount of mass as it approaches pericentre, decreasing in mass by $\sim20\%$ over the last two snapshots as it moves from $r/R_{200\rho_c}=0.65$ to its current position of $r/R_{200\rho_c}=0.41$. For comparison, the configuration-space based finder \textsc{ahf} shrinks the object by $\sim2$ over the same period. 
\label{sec:appendix:orbits}

\section{Tables}
\label{sec:appendix:tables}
We list the complete set of configuration options along with a list of properties calculated by \textsc{VELOCIraptor}.
\onecolumn
{
\centering\footnotesize
\begin{longtable}{@{\extracolsep{\fill}}p{0.125\textwidth}|p{0.2\textwidth}|c|p{0.5\textwidth}}
\caption{\textsc{VELOCIraptor} configuration parameters}
\label{tab:velociraptor:config}\\
\hline
\hline
    & Name & Recommended & Comments\\
    & & Value & \\
    \hline
    General Search Options & & & Related to general structure finding. \\
    \hline
    & Particle\_search\_type & 1
        & Integer flag setting particle types to search. All [1], Dark Matter [2], Gas [3], Star [4].\\
    & Search\_for\_substructure & 1
        & Integer flag whether to search for substructure. Yes/No [1/0].\\
    & Halo\_core\_search & 2
        & Integer flag whether to search for cores. Off [0], Flag if cores found [1], Find and grow cores [2].\\
    & Unbind\_flag & 1
        & Integer flag whether to applying an unbinding process (to substructures). Yes/No [1/0].\\
    & Bound\_halos & 0
        & Integer flag whether to applying an unbinding process to halos. Yes/No [1/0].\\
    & Baryon\_searchflag & 2 
        & Integer flag setting how baryons are treated, specifics dependent on Particle\_search\_type. No special behaviour [0]; Do not search baryons initially when searching for substructure but assign to nearest dark matter substructures in phase-space, works with All particle or DM particle search, baryons [1]; Baryons are also not used to generate links when searching for field FOF objects and also assigned to nearest dark matter substructures in phase-space, works with All particles searched [2]. DM+Baryons mode is Particle\_search\_type=1 + Baryon\_searchflag=2. \\
    & Minimum\_size & 20
        & Minimum number of particles a (sub)halo must be composed of. \\
    & Minimum\_halo\_size & -1
        & Minimum number of particles a halo must be composed of, allowing for different halo/subhalo minimum sizes. -1 Sets this to Minimum\_size. \\
    & Effective\_resolution & -1
        & For multi-resolution cosmological simulations, can set the effective resolution and there by set the inter-particle spacing (used to scale linking lengths). Code must be compiled for zoom simulations. -1 means the masses of dark matter particles are used to estimate the inter-particle spacing. \\
    & Singlehalo\_search & 0
        & Integer flag that indicates input consists of a single halo and no field halo search is required. Yes/No [0/1].\\

    \hline
    Halo Search Options & & & Related to field halo search. \\
    \hline
    & FoF\_Field\_search\_type & 3
        & Integer flag setting the field FOF algorithm to use. Adaptive 6DFOF [3], 3DFOF [5], Uniform 6DFOF (single velocity scale for all halos) [4].\\
    & Keep\_FoF & 0
        & Integer flag setting whether to keep the 3DFOF envelop of 6DFOF structures, useful for extracting stellar halos. Yes/No [1/0].\\
    & Halo\_linking\_length\_factor & 2.0
        & Factor by which the linking substructure linking length, Physical\_linking\_length, is multiplied by to find halos. \\
    & Halo\_6D\_linking\_length& & \\
    & \_factor & 1.0 
        & Factor by which the initial physical linking length of a 3DFOF halo is multiplied by when running a 6DFOF. Useful for galaxy searches.  \\
    & Halo\_6D\_vel\_linking & & \\
    & \_length\_factor & 1.25
        & Factor by which the dispersion of a 3DFOF halo is multiplied by when running a 6DFOF, $\alpha_v$ to give linking length. See \Eqref{eqn:fof6d}.  \\

    \hline
    Substructure Search Options & & & Related to velocity outlier substructure search.\\
    \hline
    & FoF\_search\_type & 1
        & Integer flag setting the substructure FOF algorithm to use. Standard phase-space algorithm [1], ROCKSTAR like core search only [6]. \\
    & Iterative\_searchflag & 1
        & Integer flag setting whether iterative search used. Yes/No [1/0].\\
    & Cell\_fraction & 0.01
        & Fraction of halo mass in each cell used to calculate background, $f_{\rm cell}$. See equations (\ref{eqn:meanpos})-(\ref{eqn:meandispvel}). \\
    & Grid\_type & 1
        & Integer flag setting type of criterion used to build mesh. Default is configuration-space shannon entropy criterion [1]; full phase-space Shannon entropy criterion [2].\\
    & Nsearch\_velocity & 32
        & Integer setting number of velocity neighbours used to calculate local velocity density $N_{\rm v}$.\\
    & Nsearch\_physical & 256
        & Integer setting number of physical neighbours to search when calculating local velocity density $N_{\rm se}$.\\
    & Outlier\_threshold & 2.5
        & Threshold to apply in phase-space algorithm, $\ELLth$. See \Eqref{eqn:linkingcriteria:ellcrit}.\\
    & Physical\_linking\_length & 0.1
        & Physical linking length. For cosmological simulations , the linking length is in units of inter-particle spacing. Otherwise, in internal units. See \Eqref{eqn:linkingcriteria:ellxcrit}.\\
    & Velocity\_ratio & 2.0
        & Velocity ratio allowed in phase-space linking. See \Eqref{eqn:linkingcriteria:vrcrit}.\\
    & Velocity\_opening\_angle & 0.18
        & Angle between velocity vectors allowed in phase-space linking. See \Eqref{eqn:linkingcriteria:thetaopcrit}.\\
    & Velocity\_linking\_length & 
        & .\\
    & Iterative\_threshold\_factor & 1.0
        & Factor multiplying $\ELLth$ when using iterative method to identify outlier regions associated with the initial candidate list of spatially compact outlier groups, $\gamma_{\mathcal{L}}$. Typical values are $\gtrsim1$. \\
    & Iterative\_linking\_length & & \\
    & \_factor & 2.0
        & Factor multiplying physical linking length when using iterative method to identify outlier regions associated with the initial candidate list of spatially compact outlier groups, $\gamma_{l_{\rm x,S}}$. Typical values are $\gtrsim1$. Common to set this value to Halo\_linking\_length\_factor, thus setting the substructure linking length to that of halos. \\
    & Iterative\_Vratio\_factor & 1.0
        & Factor multiplying velocity ratio when using iterative method to identify outlier regions associated with the initial candidate list of spatially compact outlier groups, $\gamma_{\mathcal{V}_r}$. Typical values are $\gtrsim1$. \\
    & Iterative\_ThetaOp\_factor & 1.0
        & Factor multiplying opening angle when using iterative method to identify outlier regions associated with the initial candidate list of spatially compact outlier groups, $\gamma_{\Theta_{\rm op}}$. Typical values are $\gtrsim1$. \\
    & Significance\_level & 1.0
        & Minimum significance level of group, $\beta_\mathcal{L}$, see \Eqref{eqn:ellsig}. \\ 

    \hline
    Core Search Options & & & Related to core/major merger search.\\
    \hline
    & Use\_adaptive\_core\_search & 0
        & Integer flag setting how linking lengths are scaled when searching for cores. Only scale the velocity linking length dispersions, useful in dark matter only simulations [0]; scale both configuration and velocity linking lengths using dispersions, useful for galaxy searches [1].\\
    & Use\_phase\_tensor\_core & & \\
    & \_growth & 2
        & Integer flag setting setting how cores are grown. Simple assignment using constant configuration and velocity space dispersion [0] (replace distance in \Eqref{eqn:coredist} with $D^2=({\bf x}_{i}-\bar{\bf x}_{j})/\sigma_x+({\bf v}_{i}-\bar{\bf v}_{j})/\sigma_v$); Calculate distance using phase-space tensor calculated with initial core via \Eqref{eqn:coredist}; Calculate distances with phase-space tensor via is \Eqref{eqn:coredist}, where the tensor is recalculated for each active core at each level [2].\\
    & Halo\_core\_ellx\_fac & 1.0
        & Initial factor by which the physical halo linking length is multiplied when starting core search, should be $\leq1$\\
    & Halo\_core\_ellv\_fac & 1.0
        & Initial factor by which the velocity halo linking length is multiplied when starting core search, should be $\leq1$.\\
    & Halo\_core\_ncellfac & 0.0005
        & Factor that sets the initial mininum number of particles a core must be composed of by $f_{\rm C}N_{\rm H}$, where $N_{\rm H}$ is the number of particles in the host being searched. \\
    & Halo\_core\_adaptive & & \\
    & \_sigma\_fac & 2
        & When running fully adaptive core search, this specifies the width of the physical linking length in configuration space dispersion, useful when searching for galaxies in hydrodynamical simulations. Typically values are $2$. \\
    & Halo\_core\_num\_loops & 10
        & Integer setting number of loops when searching for cores, $\Delta_{\rm C}$.\\
    & Halo\_core\_loop\_ellx\_fac & 0.8
        & Factor by which the physical linking length is multiplied at each loop when searching for cores, $\alpha_{\rm x, C}<1$, see \Eqref{eqn:coreloop}.\\
    & Halo\_core\_loop\_ellv\_fac & 1.0
        & Factor by which the velocity linking length is multiplied at each loop when searching for cores, $\alpha_{\rm v, C}<1$, see \Eqref{eqn:coreloop}.\\
    & Halo\_core\_loop\_elln\_fac & 1.2
        & Factor by which the minimum number of particles for an active core is multiplied, $\alpha_{\rm N,C}$. \\
    & Halo\_core\_phase & & \\
    & \_significance & 2.0
        & Significance a core must be in terms of phase-space distance scaled by dispersions, $\beta_{\rm C}$, see \Eqref{eqn:coresig}. \\

    \hline
    Unbinding Options & & & Related to processing candidate groups with unbinding routines.\\
    \hline
    & Unbinding\_type & 1
        & Integer flag setting the unbinding criteria that removes particles considered ``unbound'' (unbound particles meet \Eqref{eqn:binding}) . Remove unbound particles [0], remove unbound particles and possibly loosely bound particles till the fraction of the system has formally bound particles (setting $\beta_{\rm E}=1$ in \Eqref{eqn:binding}) [1]. \\
    & Allowed\_kinetic\_potential & & \\
    & \_ratio & 0.95
        & Ratio of kinetic to potential energy at which a particle is still considered bound, $\beta_{\rm E}$, see \Eqref{eqn:binding}. Values of $\beta_{\rm E}\gtrsim0.95$ keeps particles that are unlikely to leave the halo within a dynamical time, $\beta_{\rm E}=1$ is the commonly used value in configuration-space finders, and $\beta_{\rm E}\lesssim0.95$ allows one to identify unbound tidal debris.\\
    & Min\_bound\_mass\_frac & 0.6
        & Fraction of formally bound particles required $f_{\rm E}$.\\
    & Softening\_length & 0
        & Set the (simple plummer) gravitational softening length. For cosmological simulations, in units of inter-particle spacing. \\
    & Keep\_background\_potential & 0
        & Integer flag setting whether one keeps the potential of unbound particles during the unbinding process when determining whether particles are unbound. As objects are treated in isolation, it is more self-consistent to ignore the potential of particles removed from a candidate group but this potential can be retained. Yes/No [0/1].\\
    & Kinetic\_reference\_frame & & \\
    & \_type & 0
        & Integer flag setting the kinetic frame when determining whether particle is bound. Use the central regions near the centre-of-mass to determine the velocity frame [0]; Use the central region around minimum of the potential to determine the velocity frame [1]. \\
    & Min\_npot\_ref & 10
        & Integer setting the minimum number of particles used to calculate the velocity frame. \\
    & Frac\_pot\_ref & 0.1
        & Fraction of closest particles (to either the centre-of-mass or minimum potential) used to calculate the velocity frame. Typical values are $\sim0.1$.\\

    \hline
    Units & & & Converts input units to internal/output units. For accuracy, units should be chosen so quantities are close to unity. \\
    \hline
    & Length\_unit & 1.0
        & Conversion to apply to input velocity units to set internal code and output units. Example: if input was in Gpc and wanted units of gigacubits, then value is 69497252252252000. \\
    & Length\_unit\_to\_kpc & 1.0 
        & Conversion applied to the output unit to convert it to a standard unit. Example: if input was in Mpc, output was in pc (Length\_unit=1e6), then this would be 1e-3. \\
    & Velocity\_unit & 1.0 
        & Conversion to apply to input velocity units to set internal code and output units. Example: if input was in km/s and wanted units of feet/forthnight, then value is 3.96850394e9. \\
    & Velocity\_to\_kms & 1.0 
        & Conversion applied to the output unit to convert it to a standard unit. Example: if input was in kpc/Gyr, output was in cm/s (Velocity\_unit=97781.3106), then this would be 1e-5. \\
    & Mass\_unit & 1.0 
        & Conversion to apply to input velocity units to set internal code and output units. Example: if input was in 1e10 solar masses and wanted units of milliounces, then value is 7.01634377e34. \\
    & Mass\_to\_solarmass & 1.0 
        & Conversion applied to the output unit to convert it to a standard unit. Example: if input was in g, output was in earth mass (Mass\_unit=1.674481e-28), then this would be 0.000003003. \\
    & Hubble\_unit & 100.0
        & The unit of Hubble flow in internal code unit and should be H0/h in the internal Length\_unit *Velocity\_unit. Example: if internal units are kpc and km/s, then should be 0.1, if Mpc and km/s, then 100. \\
    & Gravity & 43.021
        & Gravitational constant in internal units, Length\_unit * (Velocity\_unit)$^2$ / Mass\_unit. Example: if internal units are kpc, km/s, and 1e10 solar masses, should be 43.0211349e3. \\
    & Mass\_value & 1.0
        & If code is compiled so as to not store mass (useful to save memory when processing N-Body cosmological simulations with uniform mass resolution), this sets the mass of all particles. \\

    \hline
    Simulation and Cosmology & & & \\
    \hline
    & Period & 1.0 
        & Period of simulation box. For some input formats (Gadget, RAMSES, HDF5), this is taken from the input file. \\
    & Scale\_factor & 1.0
        & Scale Factor of simulation box. For some input formats (Gadget, RAMSES, HDF5), this is taken from the input file. \\
    & h\_val & 1.0
        & The so-called little h, where the Hubble constant is h*Hubble\_unit. For some input formats (Gadget, RAMSES, HDF5), this is taken from the input file. \\
    & Omega\_m & 1.0
        & Cosmological matter density $\Omega_m$ at $a=1$. For some input formats (Gadget, RAMSES, HDF5), this is taken from the input file. \\
    & Omega\_cdm & 1.0 
        & Cosmological cold dark matter density $\Omega_{cdm}$ at $a=1$. For some input formats (Gadget, RAMSES, HDF5), this is taken from the input file. \\
    & Omega\_b & 0.0 
        & Cosmological baryon density $\Omega_{b}$ at $a=1$. For some input formats (Gadget, RAMSES, HDF5), this is taken from the input file. \\
    & w\_of\_DE & -1.0
        & Dark energy equation of state. Not yet implemented. \\
    & Omega\_Lambda & 1.0 
        & Cosmological dark energy density $\Omega_\Lambda$ (or more generally $\Omega_{\rm DE}$ at $a=1$. For some input formats (Gadget, RAMSES, HDF5), this is taken from the input file. \\
    & Critical\_density & 1.0 
        & Cosmological critical density $\rho_c=3H^2/8\pi G$ at $a=1$. For some input formats (Gadget, RAMSES, HDF5), this is taken from the input file. \\
    & Virial\_density & 500.0 
        & User defined virial overdensity $\Delta$ used to calculate the virial mass. If -1, then output virial masses will refer to the \cite{bryannorman1998} overdensity mass. \\

    \hline
    Output Options & & & \\
    \hline
    & Snapshot\_value & 0
        & Set if halo IDs should be temporally unique, useful for halo merger tree codes and analysing multiple snapshots. The resulting IDs are then the index of a halo +1 + Snapshot\_value*TEMPORALHALOIDVAL, where TEMPORALHALOIDVAL=$10^{12}$ when the code is compiled to use long integers, TEMPORALHALOIDVAL=1000000 when the code is compiled to use only integers (useful for small simulations and if worried about memory). \\
    & Inclusive\_halo\_masses & 2 
        & How masses of halos are calculated. Substructure masses are always calculated using particles exclusively belonging to the object but halos can have masses calculated in several different fashions. Use only particles belonging to the object, giving no abrupt change in spherical overdensity masses as a halo transitions from a halo to a subhalo, though overdensity masses now are not full spherical overdensity masses [0]; Use all particles in the FOF envelop, that is include the contribution from substructures [1]; Use all particles centred on the centre-of-mass, regardless of whether they belong to the halo FOF, background or another halo when calculating spherical overdensity masses [2]. \\ 
    & Comoving\_units & 0 
        & Integer flag indicating whether the properties output is in physical or comoving little h units. Yes/No [1/0]. \\ 
    & Binary\_output & 2 
        & Integer flag setting output format. HDF [2], binary [1] or ascii [0]. \\ 
    & Write\_group\_array\_file & 0 
        & Integer flag indicating whether to write a single large tipsy style group assignment file that list the group id of every particle. Yes/No [0/1]. \\ 
    & Separate\_output\_files & 0 
        & Integer flag indicating whether separate files are written for field and subhalo groups. Yes/No [0/1].\\ 
    & Extensive\_halo\_properties & & \\ 
    & \_output & 0
        & Integer flag setting whether to calculate/output even more halo properties. Yes/No [0/1]. \\ 
    & Extended\_output & 0 
        & Integer flag indicating whether to produce extended output for quick particle extraction of particles in groups from input file. Requires more memory as particles store input file and index in the file at which they are located. Yes/No [0/1]. \\ 
    & Spherical\_overdensity\_halo & & \\
    & \_particle\_list\_output & 0
        & Output list of particles in spherical overdensity regions of halos. Yes/No [0/1]. \\ 

    \hline
    Input Options & & & \\
    \hline
    & Cosmological\_input & 1
        & Integer flag indicating input data is from a cosmological simulation. Code uses cosmological information. Yes/No [0/1]. \\
    & Input\_chunk\_size & 100000
        & Amount of information to read from input file in one go, useful for managing memory when reading input data. \\ 
    & NSPH\_extra\_blocks & 0
        & Integer setting the number of extra gas/SPH particle related data blocks are to be read/are present in the file if loading gadget snapshot.\\ 
    & NStar\_extra\_blocks & 0
        & Integer setting the number of extra star particle related data blocks are to be read/are present in the file if loading gadget snapshot.\\ 
    & NBH\_extra\_blocks & 0
        & Integer setting the number of extra black hole/sink particle related data blocks are to be read/are present in the file if loading gadget snapshot.\\ 

    & HDF\_name\_convention & 1
        & Integer setting the HDF naming convention to use. Currently implemented conventions are for EAGLE, ILLUSTRIS, GIZMO/SIMBA. \\
    & Input\_includes\_star & & \\
    & \_particle & 1
        & Integer flag indicating star particles are present in the input data file. Yes/No [0/1] \\ 
    & Input\_includes\_bh & & \\
    & \_particle & 1
        & Integer flag indicating black hole/sink particles are present in the input data file. Yes/No [0/1] \\ 
    & Input\_includes\_wind & & \\
    & \_particle & 1 
        & Integer flag indicating wind particles are present in the input data file. Yes/No [0/1] \\ 
    & Input\_includes\_tracer & & \\
    & \_particle & 1 
        & Integer flag indicating tracer particles are present in the input data file. Yes/No [0/1] \\ 
    & Input\_includes\_extradm & & \\
    & \_particle & 1
        & Integer flag indicating extra dark matter types particles are present in the input data file. Yes/No [0/1] \\ 

    \hline
    Additional Options & & & \\
    \hline
    & Verbose & 0
        & Integer flag on how verbose code is during operation. Minimal [0]; Moderate [1]; Very Verbose [2].\\
    & MPI\_particle\_total & & \\
    & \_buf\_size & -1  
        & Total memory size in bytes used to store particles in temporary buffer such that particles are sent to non-reading mpi processes in one communication round in chunks of size MPI\_particle\_total\_buf\_size/Number of MPI process/memory to store a particle. Ensures that communications exceed the memory available in the MPI\_COMM\_WORLD. If -1, code determines this amount, may not be optimal. \\
    & MPI\_part\_allocation\_fac & 0.05
        & Memory allocated in MPI mode to store particles is (1+MPI\_part\_allocation\_fac)*memory to store all particles. This factor should be $\gtrsim0$ to allow room for particles to be exchanged between MPI threads without requiring new memory to be allocated. \\
    \hline
\end{longtable}
}
\twocolumn

\onecolumn
{
\centering\footnotesize
\begin{longtable}{@{\extracolsep{\fill}}p{0.2\textwidth}|p{0.175\textwidth}|p{0.6\textwidth}}
\caption{\textsc{VELOCIraptor} Outputted halo/galaxy properties.}
\label{tab:velociraptor:properties}\\
\hline
\hline
    & Name & Comments\\
    \hline
    ID and Type information & \\
    \hline
    & ID & Halo ID. ID = index of halo + 1 + TEMPORALHALOIDVAL*Snapshot\_value, giving a temporally unique halo id that can be quickly parsed for an index and a snapshot number. \\
    & ID\_mbp & Particle ID of the most bound particle in the group. \\
    & hostHaloID & ID of the host field halo. If an object is a field halo, this is -1. \\
    & Structuretype & Structure types contain information on how the object was found and at what level in the subhalo hierarchy. Field halos are 10. Substructures identified using the local velocity field are type 10+10=20, substructures identified using cores are type 10+5=15. For structures found at level 2 (ie: subhalos within subhalos), the type offset is 20, and so on. \\
    & numSubStruct & Number of substructures. Subhalos can have subsubhalos. \\
    
    \hline
    Mass and radius properties & & All properties are in output units. \\
    \hline
    & npart & Number of particles belonging exclusively to the object. \\
    & Mass\_tot & Total mass of particles belonging exclusively to the object, $M_{\rm tot}$. \\
    & Mass\_FOF & Total mass of particles in the FOF, $M_{\rm FOF}$. Is zero for substructure. \\
    & Mass\_200mean & Overdensity mass defined by the mean matter density, $M_{200\rho_m}$. For field halos, if inclusive masses are desired, this is based on the particles in the FOF. If full spherical overdensity masses are desired, then includes all particles (whether they belong to the object, the background or another object) within a spherical region. For subhalos, this is based on particles belonging exclusively to the object. \\
    & Mass\_200crit & Overdensity mass defined by the critical density, $M_{200\rho_c}$. Behaviour like Mass\_200mean. \\
    & Mass\_BN98 & Overdensity mass defined by the mean matter density and $\Delta(z)$ given by \cite{bryannorman1998}, $M_{\Delta(z)\rho_c}$. Behaviour like Mass\_200mean. \\
    & Mvir & User defined virial mass, $M_{\rm vir}$. Behaviour like Mass\_200mean. \\
    & R\_size & Maximum distance of particles belonging exclusively to the object and the centre-of-mass. \\
    & R\_200mean & Radius related to overdensity mass Mass\_200mean. \\
    & R\_200crit & \ditto\\
    & R\_BN98 & \ditto\\
    & Rvir & \ditto\\
    & R\_HalfMass & Half mass radius based on the Mass\_tot. \\

    \hline
    Position and Velocity & & All properties are in output units. Objects need not have positions periodically wrapped.  \\
    \hline
    & Xc & $x$ coordinate of centre-of-mass. \\
    & Yc & \ditto\\
    & Zc & \ditto\\
    & Xcmbp & $x$ coordinate of most bound particle. \\
    & Ycmbp & \ditto\\
    & Zcmbp & \ditto\\
    & VXc & $v_x$ velocity of centre-of-mass. \\
    & VYc & \ditto\\
    & VZc & \ditto\\
    & VXcmbp & $v_x$ velocity of most bound particle. \\
    & VYcmbp & \ditto\\
    & VZcmbp & \ditto\\

    \hline
    Velocity and Angular Momentum & & All properties are in output units. \\
    \hline
    & Vmax & Maximum circular velocity based on particles belonging exclusively to the object, where circular velocities are defined by $V_{\rm circ}^2=GM/R$. \\
    & Rmax & Radius of maximum circular velocity. \\
    & sigV & Velocity dispersion based on the velocity dispersion tensor $\sigma_v=|\Sigma|^{1/6}$, where $\Sigma$ is the velocity dispersion tensor. \\
    & veldisp\_xx & The $x,x$ component of the velocity dispersion tensor. \\
    & veldisp\_xy & \ditto\\
    & veldisp\_xz & \ditto\\
    & veldisp\_yx & \ditto\\
    & veldisp\_yy & \ditto\\
    & veldisp\_yz & \ditto\\
    & veldisp\_zx & \ditto\\
    & veldisp\_zy & \ditto\\
    & veldisp\_zz & \ditto\\
    & Lx & $x$ component of the total angular momentum about the centre-of-mass using particles belonging exclusively to the object. \\
    & Ly & \ditto\\
    & Lz & \ditto\\
    & lambda\_B & \cite{bullock2001} spin parameter $\lambda_B=$ using total angular momentum and the spherical overdensity mass, $\lambda_B=\frac{J}{\sqrt{2}MVR}$. \\
    & Krot & Measure of rotational support about the angular momentum axis $\kappa_{\rm rot}=\frac{\sum_i 1/2 m_i j_{z,i}r_i}\sum_i T_i$, where the first sum is over the motion of particles along the angular momentum axis and the second sum is over kinetic energies \cite[see][]{sales2010a}. \\

    \hline
    Morphology & & All properties are in output units. \\
    \hline

    & cNFW & Calculated assuming an NFW profile \cite[][]{nfw} following \cite{prada2012a} where we solve $\frac{V_{\rm max}^2}{GM_\Delta/R_\Delta}-\frac{0.216c}{\ln(1+c)-c/(1+c)}=0.$ \\
    & q & We calculate the shape using the reduced inertia tensor \cite[][]{dubinski1991,allgood2006}, $\tilde{I}_{j,k}=\sum\limits_n \frac{m_n x^\prime_{j,n} x^\prime_{k,n}}{(r^\prime_{n})^2}$, where the sum is over particles exclusively belonging to the object and, $(r^\prime_n)^2=(x^\prime_n)^2+(y^\prime_n/q)^2+(z^\prime_n/s)^2$ is the ellipsoidal distance between the halo's centre-of-mass and the $n$th particle, primed coordinates are in the eigenvector frame of the reduced inertia tensor and $q$ \& $s$ are the semi-major and minor axis ratios respectively. Thus q is the semi-major axis ratio. \\
    & s & \\
    & eig\_xx & \ditto\\
    & eig\_xy & \ditto\\
    & eig\_xz & \ditto\\
    & eig\_yx & \ditto\\
    & eig\_yy & \ditto\\
    & eig\_yz & \ditto\\
    & eig\_zx & \ditto\\
    & eig\_zy & \ditto\\
    & eig\_zz & \ditto\\

    \hline
    Energy & & All properties are in output units. \\
    \hline
    & Ekin & The total kinetic energy, $\sum T_i$. \\
    & Epot & The total gravitational potential energy $1/2\sum W_i$, where the 1/2 comes from double counting.\\
    & Efrac & The fraction of particles that are formally bound (i.e., have $W_i+T_i<0$.\\ 

    \hline
    Quantities within $R(V_{\rm max})$ & & Variety of properties based on particles within $r\leq R(V_{\rm max})$. \\
    \hline
    & RVmax\_sigV & Dispersion, like sigV for $r\leq R(V_{\rm max})$. \\
    & RVmax\_veldisp\_xx & Dispersion tensor, like veldisp\_xx for $r\leq R(V_{\rm max})$. \\
    & RVmax\_veldisp\_xy & \ditto\\
    & RVmax\_veldisp\_xz & \ditto\\
    & RVmax\_veldisp\_yx & \ditto\\
    & RVmax\_veldisp\_yy & \ditto\\
    & RVmax\_veldisp\_yz & \ditto\\
    & RVmax\_veldisp\_zx & \ditto\\
    & RVmax\_veldisp\_zy & \ditto\\
    & RVmax\_veldisp\_zz & \ditto\\
    & RVmax\_lambda\_B & Spin parameter, like lambda\_B for $r\leq R(V_{\rm max})$. \\
    & RVmax\_Lx & Total angular momentum, like Lx $r\leq R(V_{\rm max})$. \\
    & RVmax\_Ly & \ditto\\
    & RVmax\_Lz & \ditto\\
    & RVmax\_q & Semi-major axis ratio, like q for $r\leq R(V_{\rm max})$. \\
    & RVmax\_s & \ditto\\
    & RVmax\_eig\_xx & Eigenvectors of morphology, like eig\_xx for $r\leq R(V_{\rm max})$. \\
    & RVmax\_eig\_xy & \ditto\\
    & RVmax\_eig\_xz & \ditto\\
    & RVmax\_eig\_yx & \ditto\\
    & RVmax\_eig\_yy & \ditto\\
    & RVmax\_eig\_yz & \ditto\\
    & RVmax\_eig\_zx & \ditto\\
    & RVmax\_eig\_zy & \ditto\\
    & RVmax\_eig\_zz & \ditto\\

    \hline
    Gas quantities & & Bulk properties of gas particles/tracers when compiled to process gas properties. Properties unique to gas are T\_gas and SFR\_gas. \\
    \hline
    & n\_gas & Number of gas particles. \\
    & M\_gas & Total gas mass $M_{\rm gas}$. \\
    & M\_gas\_Rvmax & Gas mass within $R(V_{\rm max})$. \\
    & M\_gas\_30kpc & Gas mass within 30 pkpc. \\
    & M\_gas\_500c & Gas mass within a spherical overdensity of $500\rho_c$. \\
    & Xc\_gas & $x$ coordinate of centre-of-mass of gas particles relative to Xc. \\
    & Yc\_gas & \ditto\\
    & Zc\_gas & \ditto\\
    & VXc\_gas & $x$ coordinate of centre-of-mass velocity of gas particles relative to VXc. \\
    & VYc\_gas & \ditto\\
    & VZc\_gas & \ditto\\
    & Efrac\_gas & Like Efrac but for gas particles only. \\
    & R\_HalfMass\_gas & Like R\_HalfMass but for gas particles only. \\
    & veldisp\_xx\_gas & Like veldisp\_xx but for gas particles only and relative to the centre-of-mass. \\
    & veldisp\_xy\_gas & \ditto\\
    & veldisp\_xz\_gas & \ditto\\
    & veldisp\_yx\_gas & \ditto\\
    & veldisp\_yy\_gas & \ditto\\
    & veldisp\_yz\_gas & \ditto\\
    & veldisp\_zx\_gas & \ditto\\
    & veldisp\_zy\_gas & \ditto\\
    & veldisp\_zz\_gas & \ditto\\
    & Lx\_gas & Like Lx but for gas particles only and relative to the centre-of-mass. \\
    & Ly\_gas & \ditto\\
    & Lz\_gas & \ditto\\
    & q\_gas & Like q but for gas particles only and relative to the centre-of-mass. \\
    & s\_gas & Like s but for gas particles only and relative to the centre-of-mass. \\
    & eig\_xx\_gas & Like eig\_xx but for gas particles only and relative to the centre-of-mass. \\
    & eig\_xy\_gas & \ditto\\
    & eig\_xz\_gas & \ditto\\
    & eig\_yx\_gas & \ditto\\
    & eig\_yy\_gas & \ditto\\
    & eig\_yz\_gas & \ditto\\
    & eig\_zx\_gas & \ditto\\
    & eig\_zy\_gas & \ditto\\
    & eig\_zz\_gas & \ditto\\
    & Krot\_gas & Like Krot but for gas particles only and relative to the centre-of-mass.\\
    & T\_gas & Average temperature of gas. \\
    & Zmet\_gas & Average metallicity of gas. \\
    & SFR\_gas & Average star formation rate of gas. \\
    
    \hline
    Star quantities & & Bulk properties of star particles when compiled to process star properties. Similar to gas properties but has \_star instead of \_ gas. For brevity, we list only quantities unique to star particles. \\
    \hline
    & tage\_gas & Average stellar age. \\
    
    \hline
    Black hole quantities & & Bulk properties of black hole particles when compiled to process black hole properties. \\
    \hline
    & n\_bh & Number of black hole particles. \\
    & Mass\_bh & Total mass of black hole particles. \\

    \hline
    Interloper particles & & If analysing multi-resolution simulations, low resolution particles are often treated as contaminants. These are bulk properties of low resolution contaminant particles. \\
    \hline
    & n\_interloper & Number of low resolution, interloper particles. \\
    & Mass\_interloper & Total mass of low resolution, interloper particles. \\
    \hline
\end{longtable}
}
\twocolumn

\section{Associated Tools}
\label{sec:appendix:examples}
\textsc{VELOCIraptor} comes with a {\sc Python-2/3} tool-kit, specifically routines to manipulate the output data produced by the various codes. Typically, these produce {\sc dict} containing {\sc numpy} arrays, allowing for quick analysis and plotting. The repositories also come with examples of producing metric plots. The codes are  {\sc Python-3} (compatible with {\sc Python-2}) and make use of {\sc numpy}, \textsc{h5py}, \textsc{scipy}, \textsc{matplotlib}, and \textsc{scikit.learn}. 

\end{appendix}

\end{document}